\documentclass[12pt]{article}

\usepackage[a4paper,left=30mm,right=20mm,top=20mm,bottom=20mm]{geometry}
\usepackage{graphics}
\usepackage{amsmath}
\usepackage{amssymb}
\usepackage{epsfig}
\usepackage{cite}
\usepackage{xcolor}
\usepackage[unicode]{hyperref}
\graphicspath{{figures/}}

\title{Towards higher-order calculations of quarkonia production with $k_T$-factorization: $P$-wave charmonia}

\author{S.P.~Baranov$^{1}$, A.V.~Lipatov$^{2,3}$, A.A.~Prokhorov$^{3}$, X.~Chen$^{4,5}$}

\begin{document}

\maketitle

\begin{center}

{\it $^{1}$P.N. Lebedev Institute of Physics, Moscow 119991, Russia}\\
{\it $^{2}$Skobeltsyn Institute of Nuclear Physics, Lomonosov Moscow State University, 119991, Moscow, Russia}\\
{\it $^{3}$Joint Institute for Nuclear Research, 141980, Dubna, Moscow region, Russia}\\
{\it $^{4}$Institute of Modern Physics, Chinese Academy of Sciences, Lanzhou 730000, China}\\
{\it $^{5}$School of Nuclear Science and Technology, University of Chinese Academy of Sciences, Beijing 100049, China}\\

\end{center}

\vspace{0.5cm}

\begin{center}

{\bf Abstract }

\end{center}

Inclusive $P$-wave charmonia production in hadronic 
collisions at high energies is discussed
in the framework of non-relativistic QCD and $k_T$-factorization formalism.
We present two consistent approches to merge the usual 
leading order $k_T$-factorization calculations with tree-level next-to-leading 
order off-shell amplitudes. 
Using these prescriptions, 
we extracted 
long-distance matrix elements for $\chi_c$ mesons 
from a combined fit to available Tevatron and LHC data.
In contrast to previous (leading order) calculations, our fits 
do not contradict equal color 
singlet wave functions of $\chi_{c1}$ and $\chi_{c2}$ states.
The extracted values of long-distance matrix elements are employed
to analyse the $\chi_c$ polarization data reported recently by the CMS Collaboration.
Our predictions are in a reasonably good agreement with the Tevatron
and LHC measurements within the theoretical and experimental uncertainties.

\indent

\vspace{1cm}


\newpage

\section{Introduction} \indent

Up to now, the production of heavy quarkonia (charmonia and bottomonia) in high energy hadronic collisions is under intense 
theoretical and experimental study\cite{Quarkonia-Review-1, Quarkonia-Review-2, Quarkonia-Review-3}.
It provides a sensitive tool probing Quantum Chromodynamics (QCD) in both perturbative and 
non-perturbative regimes, as the production mechanism involves both short and 
long distance interactions.
A rigorous framework for the description of heavy quarkonia production 
is the non-relativistic QCD (NRQCD)\cite{NRQCD-1, NRQCD-2},
which is based on a double series expansion of perturbation theory in the 
strong coupling $\alpha_s$ and the relative velocity of quarks $v$.
In this way, the perturbatively calculated cross sections for the short distance 
production of a heavy quark pair $Q\bar Q$ in an intermediate Fock state $^{2S + 1}L_J^{[a]}$ with 
spin $S$, orbital angular momentum $L$, total angular momentum $J$ and color representation $a$ 
(singlet, $a = 1$, or octet, $a = 8$) are accompanied with long distance matrix elements (LDMEs) 
which describe the subsequent non-perturbative transition of the
intermediate $Q\bar Q$ pair into a physical meson via soft gluon radiation.
Treating the soft transition probabilities as free parameters in the framework of collinear
factorization approach at the next-to-leading order (NLO), a good description has been achieved 
for the charmonia ($J/\psi$, $\psi^\prime$, $\chi_c$) and bottomonia ($\Upsilon$, $\chi_b$) 
transverse momentum distributions (see, for example,\cite{Charmonia-NRQCD-1, Charmonia-NRQCD-2, 
Charmonia-NRQCD-3, Charmonia-NRQCD-4, Charmonia-NRQCD-5, Charmonia-NRQCD-6, Charmonia-NRQCD-7} 
and\cite{Bottomonia-NRQCD-1, Bottomonia-NRQCD-2, Bottomonia-NRQCD-3, Bottomonia-NRQCD-4, 
Bottomonia-NRQCD-5, Bottomonia-NRQCD-6}, respectively).
A possible solution to a long-standing problem known in the literature as the 
"Polarization Puzzle"\cite{PolarizationPuzzle-1, PolarizationPuzzle-2, PolarizationPuzzle-3} 
and the "Heavy Quark Spin Symmetry Puzzle"\cite{HQSSPuzzle-1, HQSSPuzzle-2} has been recently
proposed\cite{TransitionMechanism}, that could lead to a consensus on the mechanism of quarkonium
formation\footnote{A competing theoretical approach, the so called Improved Color Evaporation 
Model\cite{CEM-1, CEM-2, CEM-3, CEM-4}, has failed to describe the LHC data on the $J/\psi$ 
production at large transverse momenta, on the double $J/\psi$ production and on the $J/\psi$ 
production in $e^+ e^-$ annihilation, see\cite{Charmonia-NRQCD-2, CEMProblems}
for more information.}.
Also, a calculation of tree-level next-to-next-to-leading order (NNLO$^*$) corrections 
to the color-singlet mechanism in the collinear scheme has become 
available\cite{Quarkonia-NNLO-1, Quarkonia-NNLO-2}.

At high energies, a large piece of tree-level NLO + NNLO + ... corrections to the perturbative
production of heavy quark pairs can be efficiently taken into account in the framework of the
$k_T$-factorization\cite{kt-factorization} (or high energy factorization\cite{HighEnergyFactorization})
approach. These corrections correspond to the diagrams with real gluon emissions in initial state, which
dominate over other possible corrections at high energies.
The $k_T$-factorization approach is based on the Balitsky-Fadin-Kuraev-Lipatov (BFKL)\cite{BFKL} or
Ciafaloni-Catani-Fiorani-Marchesini (CCFM)\cite{CCFM} evolution equations for the gluon densities
in the proton. The latter are known as the Transverse Momentum Dependent (TMD) or unintegrated gluon 
densities. This method can be considered as a convenient alternative to explicit high-order pQCD 
calculations. A detailed description and discussion of the $k_T$-factorization technique can be found
in the review\cite{TMD-Review}.
Nowadays, it has become a widely exploited tool and, being supplemented with the NRQCD formalism,
has been successfully applied to the charmonia and bottomonia production at modern colliders (see for 
example \cite{Our-Charmonia-1, Our-Charmonia-2, Our-Charmonia-3, Our-Bottomonia-1, Our-Bottomonia-2, 
Our-Bottomonia-3} and references therein). A good agreement has been obtained with the LHC data, 
including the polarization observables for $J/\psi$, $\psi^\prime$ and $\Upsilon(nS)$ mesons.

However, the $k_T$-factorization fits\footnote{Based on leading order ${\cal O}(\alpha_s^2)$ production amplitudes, see below.} to the experimental data 
require unequal values of the color singlet wave functions
for $P$-wave states $\chi_{c1}$ and $\chi_{c2}$. The values\cite{Our-Charmonia-1, Our-Charmonia-3}
extracted from the LHC measurements may differ from each other by a factor of $2$ --- $4$
(see also\cite{UnEqualWaveFunctions}). 
Analogous results have been also found for some $P$-wave bottomonium states, namely, $\chi_{b1}(1P)$ 
and $\chi_{b2}(1P)$ mesons\cite{Our-Bottomonia-3}. These results are at variance with the Heavy Quark
Spin Symmetry (HQSS) relations\cite{NRQCD-1, NRQCD-2}, which are valid up to ${\cal O}(v^2)$ accuracy.
According to the HQSS relations, the color singlet wave functions and/or LDMEs for different total angular momentum $J$ can
only differ by an overall normalization factor representing the averaging over the spin degrees of freedom.
The situation awaits for an explanation.

On the one hand, 
one can argue\cite{UnEqualWaveFunctions, Our-Charmonia-1, Our-Charmonia-3} that treating 
the charmed quarks as spinless particles in the potential models might be an oversimplification
(see \cite{PotentialModelCalcutations-1, PotentialModelCalcutations-2}), 
or that the radiative corrections to the wave functions may be large (as they are known to be
for $J/\psi$ meson).
But, on the other hand, the inconsistency may come from the fact that the up-to-now calculations%
\cite{Our-Charmonia-1, Our-Charmonia-2, Our-Charmonia-3, Our-Bottomonia-1, Our-Bottomonia-2, 
Our-Bottomonia-3, UnEqualWaveFunctions} were only limited to the leading order in $\alpha_s$.
Thus, one can hope that after taking into account additional higher-order contributions 
(not encoded in the CCFM-evolved TMD gluon densities) the HQSS relations could be restored.
So, the main goal of the present study is to perform the NLO calculations in the $k_T$-factorization
approach and to test the corresponding predictions at the Tevatron and LHC conditions\footnote{The necessity of 
NLO NRQCD terms to explain the available data on $\chi_c$ production within the collinear QCD factorization was pointed out\cite{Charmonia-NRQCD-7}.}.

A well known difficulty in this kind of calculations is to properly avoid double counting.
Indeed, the same gluon emission act can be accounted twice: as a part of the initial state 
radiation cascade (which is described as the evolution of TMD parton density) and as a part 
of the hard interaction process (which is described as an explicit NLO contribution). 
For more discussion see the review \cite{TMD-Review} and references therein.
To avoid this double counting one has to properly match the LO and NLO off-shell amplitudes.
Below we will adopt for our purposes a prescription which has been consistently applied to 
the $c$-jet production\cite{MatchingProcedure-1} and to the associated $W^\pm$ or $Z$ and 
heavy quark production\cite{MatchingProcedure-2} at the LHC.
Such calculations for heavy quarkonia will be performed for the first time.

The paper is organized as follows. In Section 2 we briefly describe
our theoretical framework and basic steps of our calculation. In Section 3
we present the numerical results and discussion.
Section~4 sums up our conclusions.

\section{Theoretical framework} \indent

This section provides a brief review of the $k_T$-factorization formulas
for $P$-wave charmonia production and a short description of the calculation
steps. 

\subsection{Basic formulas} \indent

\begin{figure}
\begin{center}
{\includegraphics[width=.45\textwidth]{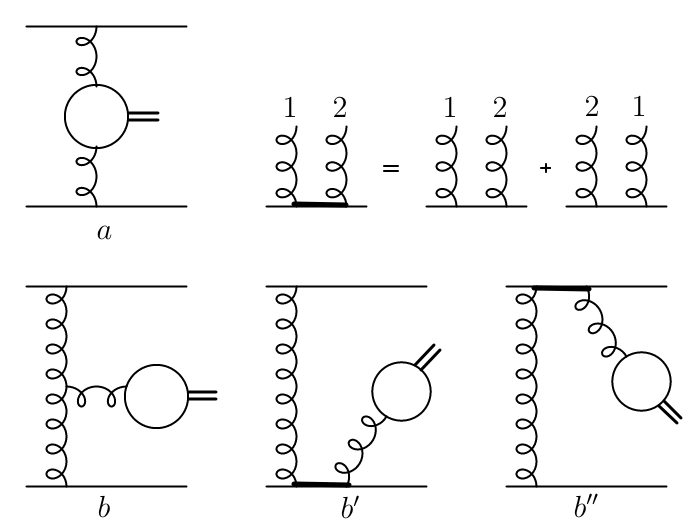}}\hfill
\caption{Feynman diagrams which represent the $2 \to 1$ contributions to charmonia production.}
\label{fig:LOdiag}
 \end{center}
\end{figure}

The true leading order (LO) contributions are represented by a number of off-shell (dependent 
on the non-zero virtualities of incoming particles)
$2\to 1$ gluon-gluon fusion subprocesses resulting in the production of a $c\bar{c}$
pair in the color singlet or color octet state. 
These processes are of ${\cal O}(\alpha_s^2)$ order, in contrast with the collinear
QCD factorization where the perturbative expansion starts from the $2\to 2$ 
subprocesses. So, we have:
\begin{gather}
  g^*(k_1)+g^*(k_2)\to c\bar c\left[ \, ^3P_J^{[1]},\, ^3S_1^{[8]},\ ^1P_1^{[8]}\right] (p),
  \label{eq:LO}
\end{gather}
\noindent
where the four-momenta of all particles are indicated in the parentheses.
Corresponding Feynman diagrams are shown in Fig.~\ref{fig:LOdiag}.
The color octet states further evolve into real mesons by non-perturbative QCD 
transitions: $c\bar c\to\chi_{cJ}+X$ with $J=0$, $1$ or $2$ where $X$ may have quantum
numbers of one or several gluons.
The last contribution in (\ref{eq:LO}) is formally suppressed by two extra powers of
the relative charmed quark velocity $v$; it was, however, argued\cite{Charmonia-NRQCD-6}
that it could still be non-negligible and has to be taken into consideration.
The calculation of $2 \to 1$ production amplitudes~(\ref{eq:LO}) is straightforward
(see, for example,\cite{Our-Charmonia-1, Our-Charmonia-2, Our-Charmonia-3} and references therein for more details).
Here we only mention that the polarization tensor of incoming off-shell
gluons is taken in the specific BFKL form\cite{kt-factorization, HighEnergyFactorization}: 
$\sum \epsilon_i^\mu \epsilon_i^{*\nu} = k_{iT}^\mu k_{iT}^\nu/{\mathbf k}_{iT}^2$,
where $k_{iT}^2 = k_i^2 = - {\mathbf k}_{iT}^2$ with $i = 1, 2$.

\begin{figure}
\begin{center}
{\includegraphics[width=.65\textwidth]{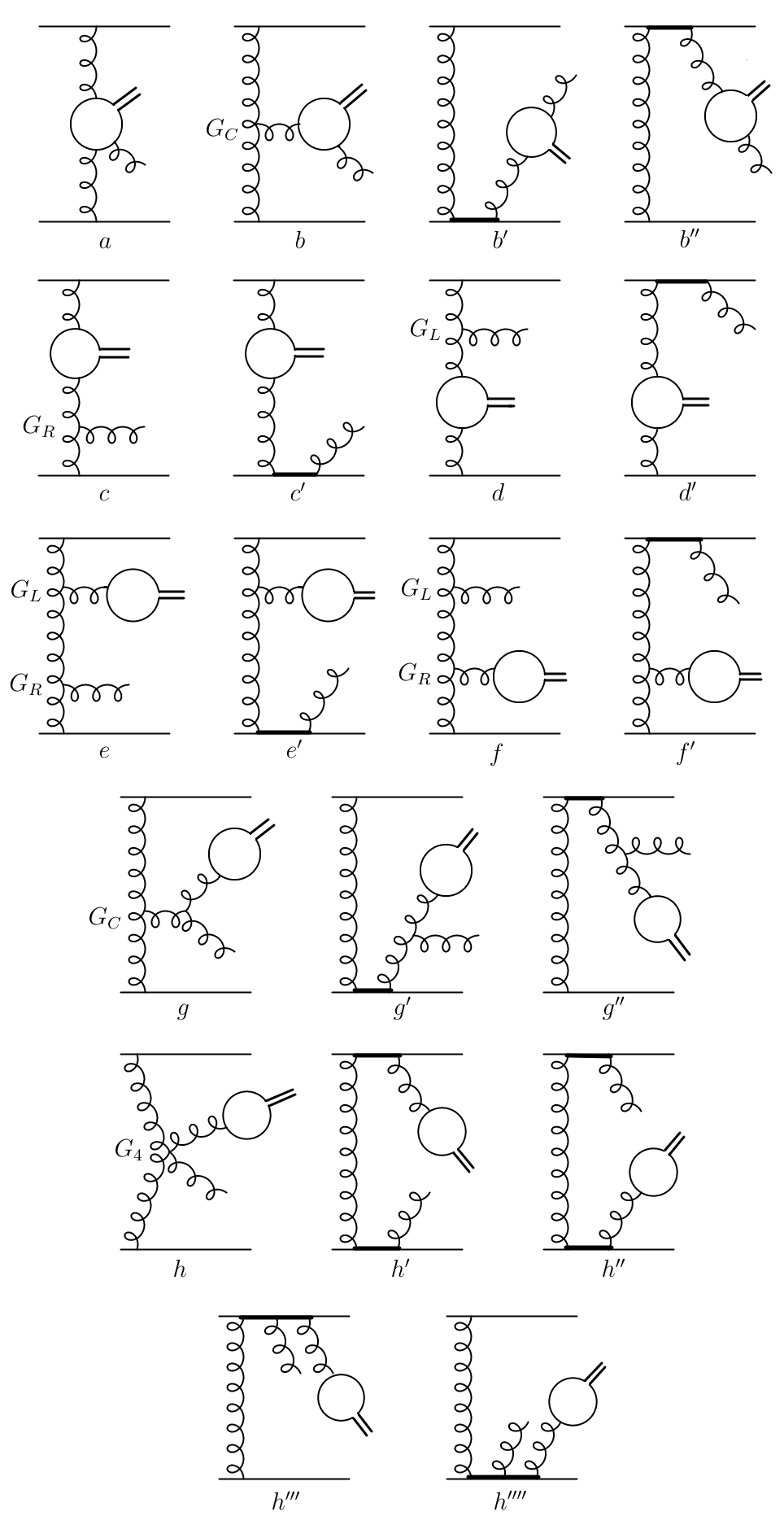}}\hfill
\caption{Feynman diagrams which represent the $2 \to 2$ contributions to charmonia production.}
\label{fig:NLOdiag}
 \end{center}
\end{figure}

The next-to-leading order (NLO) is represented by the ${\cal O}(\alpha_s^3)$ 
subprocesses shown in Fig.~\ref{fig:NLOdiag}:
\begin{gather}
  g^*(k_1)+g^*(k_2)\to c\bar c\left[ \, ^3P_J^{[1]},\, ^3S_1^{[8]},\ ^1P_1^{[8]} \right] (p)+g(p_g).
  \label{eq:NLO}
\end{gather}
\noindent
The evaluation of Feynman diagrams was partly described\cite{SPB-Quarkonia-22}.
An essential point in doing the calculations is that the initial gluon off-shellness
may violate gauge invariance. To solve this problem, we follow the technique\cite{EffectiveAction-1, EffectiveAction-2, EffectiveAction-3}. 
First, we start with an extended set of diagrams 
where the off-shell gluon lines are considered as internal lines emitted by external
quark fields, while the external quark fields are on-shell. 
Then, we apply the eikonal approximation for the emission of gluons, and this allows
us to absorb the contributions from non-factorizable diagrams into factorizable ones,
by means of modifying the expressions for three- and four-gluon couplings\footnote{Except case of four-gluon coupling, one can 
alternatively use the BFKL form of gluon polarization tensor.}. Namely, 
the diagrams of the type $b'$ and  $b''$ can be absorbed into the diagrams of the type
$b$; the diagrams of the type $c'$ and $d'$ can be absorbed 
into the diagrams of the type $c$ and $d$, respectively, and so on.
The explicit expressions for the
modified three- and four-gluon vertices $G_L$, $G_R$, $G_C$ and $G_4$ are presented\cite{EffectiveVertices}. 
So, for the gluons having the momenta $k_1$, $k_2$, $k_3$, $k_4$ and the respective
colors $a$, $b$, $c$, $d$ these vertices read:
\begin{gather}
  G_{abc}^{\mu\nu\lambda}(k_1,k_2,k_3) = g f_{abc}\left[(k_2^\lambda - k_1^\lambda) g^{\mu\nu} + (k_3^\mu - k_2^\mu) g^{\nu\lambda} + (k_1^\nu - k_3^\nu) g^{\lambda \mu} \right], \\
  G_{L\,abc}^{\epsilon_1\nu\lambda}(k_1,k_2,k_3) = G_{abc}^{\epsilon_1\nu\lambda}(k_1,k_2,k_3) - g f_{abc} {\epsilon_1^\nu \epsilon_1^\lambda k_1^2 \over (\epsilon_1 k_3)}, \\
  G_{R\,abc}^{\mu\epsilon_2\lambda}(k_1,k_2,k_3) = G_{abc}^{\mu\epsilon_2\lambda}(k_1,k_2,k_3) - g f_{abc} {\epsilon_2^\mu \epsilon_2^\lambda k_2^2 \over (\epsilon_2 k_3)}, \\
  G_{C\,abc}^{\epsilon_1\epsilon_2\lambda}(k_1,k_2,k_3) = G_{abc}^{\epsilon_1\epsilon_2\lambda}(k_1,k_2,k_3) + 2g f_{abc} \left[{\epsilon_1^\lambda k_1^2 \over (\epsilon_1 k_2)} - {\epsilon_2^\lambda k_2^2 \over (\epsilon_2 k_1}) \right],\\
  G_{4\,abcd}^{\epsilon_1\epsilon_2\lambda\sigma}(k_1,k_2,k_3,k_4) = 
      ig^2 f_{abr}f_{cdr}\left[\epsilon_1^\sigma \epsilon_2^\lambda - \epsilon_1^\lambda \epsilon_2^\sigma\right] + \nonumber \\
    + ig^2 f_{adr}f_{bcr}\left[\epsilon_1^\lambda\epsilon_2^\sigma - (\epsilon_1 \epsilon_2) g^{\lambda \sigma} -\frac{2k_2^2\epsilon_2^\sigma \epsilon_2^\lambda}{(\epsilon_2 k_3)(\epsilon_2 k_1)} - \frac{2k_1^2 \epsilon_1^\sigma \epsilon_1^\lambda}{(\epsilon_1 k_4)(\epsilon_1 k_2)}\right] + \nonumber \\
    + ig^2 f_{acr}f_{dbr}\left[(\epsilon_1 \epsilon_2) g^{\lambda \sigma} - \epsilon_1^\sigma \epsilon_2^\lambda - \frac{2k_1^2 \epsilon_1^\sigma \epsilon_1^\lambda}{(\epsilon_1 k_3)(\epsilon_1 k_2)} - \frac{2k_2^2 \epsilon_2^\sigma \epsilon_2^\lambda}{(\epsilon_2 k_4)(\epsilon_2 k_1)}\right],
\end{gather}
\noindent
where $f_{abc}$ are the SU$(3)$ structure constants.
The quantities
\begin{gather}
 \epsilon_i^\mu =  p_i^\mu x_i/|{\mathbf k}_{iT}|
 \label{eq:PolarizationVector}
\end{gather}
\noindent
play the role of incoming gluon polarization vectors;
here $p_i^\mu$ are the initial proton momenta, $x_i$ are the gluon longitudinal momentum
fractions and $k_{iT}$ are the (non-zero) gluon transverse momenta.
The above effective vertices ensure the gauge invariance of the whole set of amplitudes 
despite the incoming gluons are off-shell.

The production amplitudes contain projection 
operators\cite{ProjectionOperators-1, ProjectionOperators-2, ProjectionOperators-3, ProjectionOperators-4, ProjectionOperators-5}
which discriminate the spin-singlet and spin-triplet $c\bar c$ states:
\begin{gather}
 \Pi_0 = {1\over (2m_c)^{3/2}} (\hat p_{\bar c} - m_c) \gamma_5 (\hat p_{c} + m_c), \nonumber \\
 \Pi_1 = {1\over (2m_c)^{3/2}} (\hat p_{\bar c} - m_c) \hat \epsilon(S_z) (\hat p_{c} + m_c),
 \label{eq:ProjectionOperators}
\end{gather}
\noindent
where $m_c$ is the charmed quark mass, $p_c$ and $p_{\bar c}$ are the charmed quark
and antiquark momenta, $p_{c} = p/2 + q$ and $p_{\bar c} = p/2 - q$. States with 
different projections of the spin momentum onto the $z$ axis are represented by the 
polarization four-vector $\epsilon(S_z)$, and the relative momentum $q$ of the quarks 
in a bound state is associated with the orbital angular momentum $L$.
According to the general formalism\cite{ProjectionOperators-1, ProjectionOperators-2, ProjectionOperators-3, ProjectionOperators-4, ProjectionOperators-5},
the terms showing no dependence on $q$ are identified with the contributions to the 
$L = 0$ states while the terms linear in $q$ are related to the $L = 1$ states with 
the polarization vector $\epsilon(L_z)$. The states with definite projections of the 
spin and orbital momenta $S_z$ and $L_z$ can be translated into states with definite
total angular momentum $J_z$ (that is, the real mesonic states $\chi_{c0}$, 
$\chi_{c1}$, $\chi_{c2}$) through Clebsch-Gordan coefficients.

The analytical expressions for $2 \to 2$ off-shell
matrix elements were obtained using the algebraic manipulation system \textsc{form}\cite{FORM}.
We have checked that in the on-shell limit we recover the well-known results\cite{OnShellAmplitudes22-Quarkonia}.

To describe non-perturbative transitions of the color-octet $c\bar c$ pairs into
real final state mesons we employ an approach\cite{TransitionMechanism} based on 
classical multipole radiation theory; the soft gluon emission amplitudes are taken
identical to the ones describing real radiative transitions 
$\psi(2S)\to\chi_J+\gamma$ or $\chi_J\to J/\psi+\gamma$.
This approach results in a good
description of the available LHC data on charmonia and bottomonia
polarizations (see\cite{Our-Charmonia-1, Our-Charmonia-2, Our-Charmonia-3, Our-Bottomonia-1,
Our-Bottomonia-2, Our-Bottomonia-3} for more details).

According to the $k_T$-factorization prescription\cite{kt-factorization, HighEnergyFactorization},
the cross section of the considered processes is calculated as a convolution of the off-shell production amplitude $\bar {|{\cal A}|^2}$ and TMD gluon densities in a proton, 
$f_g(x, {\mathbf k}_T^2, \mu^2)$.
Thus, the cross sections for the $2 \to 1$ and $2 \to 2$ subprocesses (\ref{eq:LO}) and (\ref{eq:NLO}) can be written as:
\begin{gather}
  \sigma_{2 \to 1} = \int {2 \pi \over x_1 x_2 s F} f_g(x_1, {\mathbf k}_{1T}^2, \mu^2) f_g(x_2, {\mathbf k}_{2T}^2, \mu^2) |{\cal \bar A}|^2_{2 \to 1} \, d{\mathbf k}_{1T}^2 d{\mathbf k}_{2T}^2 dy {d\phi_1 \over 2\pi} {d\phi_2 \over 2\pi}, \\
  \sigma_{2 \to 2} = \int {1 \over 8\pi (x_1 x_2 s) F} f_g(x_1, {\mathbf k}_{1T}^2, \mu^2) f_g(x_2, {\mathbf k}_{2T}^2, \mu^2) |{\cal \bar A}|^2_{2\to 2} \times \nonumber \\ 
  \times d{\mathbf p}_{T}^2 d{\mathbf k}_{1T}^2 d{\mathbf k}_{2T}^2 dy dy_g {d\phi_1 \over 2\pi} {d\phi_2 \over 2\pi},
\end{gather}
\noindent
where $\phi_1$ and $\phi_2$ are the azimuthal angles of the initial off-shell gluons, 
$p_T$ and $y$ are the transverse momentum and rapidity of the produced $\chi_c$ meson, 
$y_g$ is the rapidity of the outgoing gluon, $\sqrt s$ is the $pp$ center-of-mass energy,
$\mu$ is the hard interaction scale and $F = 2\lambda^{1/2}(\hat s, k_1^2, k_2^2)$ is the flux 
factor\footnote{In the case of $2 \to 2$ processes, one can use the standard expression 
$\lambda^{1/2}(\hat s, k_1^2, k_2^2) \simeq x_1 x_2 s$,
see discussion\cite{Flux-2} for more details.}, where $\hat s = (k_1 + k_2)^2$\cite{Flux-1}.
The necessary matching procedure for $2 \to 1$ and $2 \to 2$ contributions is discussed below.

\subsection{TMD gluon densities in a proton} \indent

For the TMD gluon densities in a proton, we have tried two recent sets, referred to as LLM'2022\cite{LLM-2022} and JH'2013 set 2\cite{JH2013}, and a rather old set A0\cite{A0}. 
All these gluon densities are obtained from a numerical solution of the CCFM equation 
(at the leading logarithmic approximation, LLA) and are widely used in phenomenological applications (see, for example,\cite{Motyka-photon, LMJ-PP, LM-Higgs, LLM-FL, LLM-photon} 
and references therein).
The parameters of (rather empirical) input distributions employed in the JH'2013 and A0 sets
were derived from a fit to the HERA data on the proton structure functions $F_2(x, Q^2)$ and
$F_2^c(x, Q^2)$ at small $x$.
An analytical expression for the seed TMD gluon density in the very recent LLM'2022 set 
was suited 
to the best description of the LHC data on the charged hadron production at low transverse momenta $p_T \sim 1$~GeV in the framework of modified soft quark-gluon string model\cite{ModifiedSoftQuarkGluonStringModel-1, ModifiedSoftQuarkGluonStringModel-2},
with taking into account the gluon saturation effects, which are important at low scales.
Some phenomenological parameters were derived from the LHC and HERA data on several hard
QCD processes (see\cite{LLM-2022} for more information). 
All these TMD gluon densities are available from the far-famed \textsc{tmdlib} package\cite{TMDLib2},
which is a C$++$ library providing a framework and an interface to the different 
parametrizations\footnote{Unfortunately, the next-to-leading logarthmic corrections to the CCFM equation are yet not known. 
However, it can be argued\cite{CASCADE2} that the CCFM evolution at the LLA leads to reasonable QCD predictions.}.

\subsection{Numerical parameters} \indent

Following\cite{PDG}, we set the meson masses to $m(\chi_{c1}) = 3.51$~GeV, 
$m(\chi_{c2}) = 3.56$~GeV, $m(J/\psi) = 3.096$~GeV and branching fractions
$B(\chi_{c1} \to J/\psi + \gamma) = 33.9$\%, $B(\chi_{c2} \to J/\psi + \gamma) = 19.2$\% 
and $B(J/\psi \to \mu^+\mu^-) = 5.961$\%. 
We use the one-loop expression for the QCD coupling $\alpha_s$ with $n_f = 4$ quark
flavours at $\Lambda_{\rm QCD} = 250$~MeV for A0 gluon density, and the
two-loop expression for $\alpha_s$ with $n_f = 4$ and $\Lambda_{\rm QCD} = 200$~MeV for LLM'2022 and JH'2013 set 2 densities. 
Our default choice for the renormalization scale $\mu_R$ is the transverse mass of the produced
meson. The factorization scale $\mu_F$ was set to $\mu_F^2 = \hat s + {\mathbf Q}_T^2$,
where ${\mathbf Q}_T$ is the net transverse momentum of the initial off-shell gluon pair.
The choice of $\mu_R$ is rather standard for charmonia production,
while the special choice of $\mu_F$ is connected with (specific for) the CCFM evolution (see\cite{CCFM}).

\subsection{Matching the $2 \to 1$ and $2 \to 2$ contributions} \indent

We now discuss the procedure for matching the $2\to 1$ and $2\to 2$ contributions, which is necessary to avoid the double counting mentioned above. 
The proper treatment is not an easy task since there is a lack of well 
established theoretical techniques.
Our approach is mainly based on a prescription apllied earlier\cite{MatchingProcedure-1, MatchingProcedure-2}.
The main idea is to include the $2 \to 2$ contributions with certain limitations, 
so that to exclude the terms already taken into account as part of the CCFM evolution
of gluon densities.
Below we consider two possible matching scenarios.

\subsubsection{Scenario A} \indent

This scenario is based on the notion that the emission of high-$p_T$ gluons is mainly 
due to hard parton interaction, while the emission of softer gluons can be included in 
the TMD gluon density.
Within the proposed scheme, we sum together the $2\to 1$ and $2\to 2$ contributions,
taking the $2\to 1$ subprocess without any restrictions, but put constraints on the 
CCFM evolution in the case of $2\to 2$ subprocess.

On the average, the transverse momenta of the gluons emitted during the 
evolution decrease from the hard interaction block
to the proton. Assuming that the gluon emitted at the last evolution step compensates
the whole transverse momentum of the gluon participating in the hard subprocess, we introduce
a double-counting-exclusion (DCE) cut: 
$|{\mathbf p}_{gT}| > \max(|{\mathbf k}_{1T}|, |{\mathbf k}_{2T}|)$.
It excludes the terms generated by the CCFM evolution (explicitly presented 
in the $2 \to 1$ contributions) and ensures that hardest gluon 
emission in the $2 \to 2$ events comes from the hard matrix element. 
The evolution scale in the $2 \to 2$ subprocesses has to be shifted to 
the produced meson transverse mass, $\mu_F^2 \to m_T^2$,
that corresponds to the standard evolution scale used in the $2 \to 1$ subprocesses.
So, in this way the leading ($2 \to 1$) and next-to-leading ($2 \to 2$) contributions 
can be consistently summed together without double counting.

\begin{figure}
\begin{center}
{\includegraphics[width=.75\textwidth]{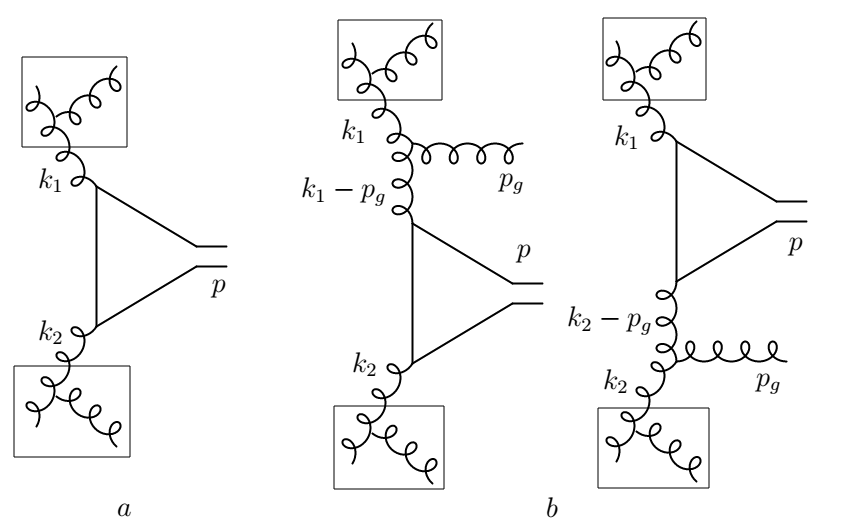}}\hfill
\caption{Feynman diagrams for $\chi_c$ production via $^3P^{[1]}_J$ intermediate state for $2\rightarrow 2$ (b) subprocess 
which partially covered by $2\rightarrow 1$ (a). The boxes represent multiple gluon emissions generated by the CCFM evolution.}
\label{fig:scenarioB_diag}
 \end{center}
\end{figure}

\subsubsection{Scenario B} \indent

This scenario is based on the observation that only certain sets of $2 \to 2$
diagrams for some terms can contribute to the double counting.
For example, the final state gluon 
emitted from the quark line is not taken into account in the terms 
generated by the CCFM evolution neither in $2 \to 2$ nor $2 \to 1$ subprocesses,
see Figs.~\ref{fig:LOdiag} and \ref{fig:NLOdiag}. It is clear that such contributions, of course, cannot be a source of the double counting but nevertheless
fall under the limitations of scenario A.
Here we try more target restrictions mainly addressed to these diagrams. 
It could allow us to avoid the double counting 
without imposing significant restrictions for the rest ones.

Let us consider $^3P_J^{[1]}$ production mechanism.
It is well known that taking the BFKL form for off-shell gluon polarization tensor 
sends to zero the contribution from non-factorizable $2 \to 2$ diagrams shown in Fig.~\ref{fig:NLOdiag}.
Only two of the remaining diagrams 
(namely, the diagrams of the type $c$ and $d$) find
themselves in the corresponding $2 \to 1$ terms
supplemented with additional gluon emissions in the CCFM evolution cascade, see Fig.~\ref{fig:scenarioB_diag}.
To avoid relevant double counting one can limit the integration over the transverse momenta of the 
incoming off-shell gluons in the $2 \to 1$ subprocess from above with some value $k_T^{\rm cut}$.
At the same time, only the events with 
minimal gluon propagator $\sqrt t = \min \{ (k_1 - p_g)^2, (k_2 - p_g)^2 \}^{1/2}$
larger than the cut scale $k_T^{\rm cut}$ are accepted when calculating the $2 \to 2$ contributions.
The exact $k_T^{\rm cut}$ value could be determined 
from the continuously merged $d\sigma_{2 \to 1}/dq_T$ and $d\sigma_{2 \to 2}/d \sqrt t$ 
distributions, where $q_T$ is the transverse momentum of any initial gluon in the $2 \to 1$ subprocess (see below).
In this way one can also almost avoid the double counting region\footnote{The similar approach 
has been used earlier\cite{MatchingProcedure-1, MatchingProcedure-2}.}.
So that, we propose the following merging scheme:
\begin{gather}
 \circ \ \ 2 \rightarrow 1: \left\{ \begin{aligned} & P = 1/2: k_{1T} < k^{\textup{cut}}_{T}, \, k_{2T} \,\,\, \textup{without cuts} \\
  										& P = 1/2: k_{1T} \,\,\, \textup{without cuts}, \, k_{2T} < k^{\textup{cut}}_{T} \end{aligned}\right.  \ \ \ \ \ \ \  \circ \ \  2 \rightarrow 2: \sqrt{t} > k^{\textup{cut}}_{T}
\end{gather} 
\noindent
where the probabilities $P = 1/2$ are 
due to the symmetry of diagrams shown in Fig.~\ref{fig:scenarioB_diag}.

Here we note that final state gluon, 
produced in the hard $2 \to 2$ subprocess, should resolve 
the charmed quark and antiquark before they form the intermediate Fock state.
Therefore, it's wavelength should be less than the typical transverse
size of the latter.
This requirement leads to a simple condition: $E_g^* > m_{c\bar c} v_c$,
where $E_g^*$ is the emitted gluon energy (in the charmonium rest frame), 
$m_{c\bar c}$ is the mass of produced $c\bar c$ state
and $v_c^2 \sim 0.23$\cite{NRQCD-1, NRQCD-2} is the relative velocity of the charmed quarks.
Condition above preserve us from the collinear divergencies
which originate when the final state gluon is emitted close to the charmed 
quark\footnote{In the scenario A, this condition is absorbed into the DCE cut.}.

Note also that the proposed scheme cannot be applied
for intermediate $^3S_1^{[8]}$ state due to a 
presence of non-factorizable diagrams with two $t$-channel gluons (type $e$ and $f$ diagrams, Fig.~\ref{fig:NLOdiag}).
So, for this case we will exploit scenario A.

\subsubsection{Determination of $k_T^{\rm cut}$ and role of NLO$^*$ corrections} \indent

\begin{figure}
\begin{center}
{\includegraphics[width=.7\textwidth]{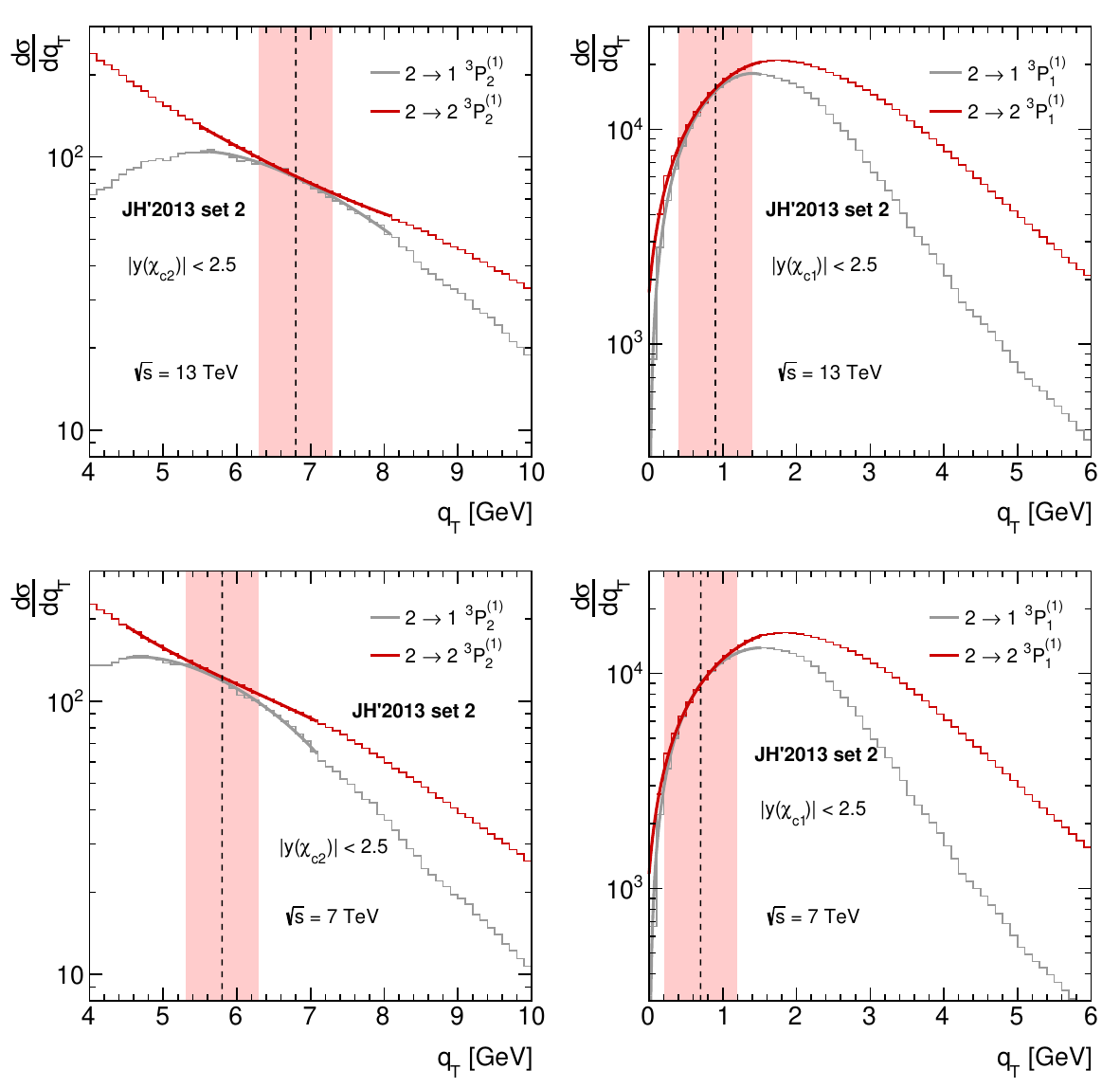}}\hfill
\caption{Matching the $d\sigma_{2\rightarrow 1}/dq_{T}$ and $d\sigma_{2\rightarrow 2}/d\sqrt{t}$ spectra 
for color singlet $^3P^{[1]}_1$ (right panel) and $^3P^{[2]}_1$ (left panel) in the central 
rapidity region $|y| < 2.5$ at $\sqrt{s} = 7$ and $13$~TeV.
The JH'2013 set 2 gluon density is used. 
Shaded pink bands represent a $\pm\,0.5$ GeV variation in the $k_T^{\rm cut}$ values.}
\label{fig:kt_cut}
 \end{center}
\end{figure}

As it was mentioned above, 
a reasonable choice for $k_T^{\rm cut}$, which is an essential part of scenario B, can be 
provided by the touch (meeting) point of the $d\sigma_{2 \to 1}/dq_T$ and 
$d\sigma_{2 \to 2}/d \sqrt t$ spectra.
We perform these calculations in the rapidity region $|y(\chi_{cJ})| < 2.5$, which is close to 
the experimental conditions of the CMS and ATLAS Collaborations at the LHC.
Our results for both color singlet states, $^3P^{[1]}_1$ and $^3P^{[1]}_2$, are shown in Fig.~\ref{fig:kt_cut},
where the dotted vertical line specifies the $k_{T}^{\rm cut}$ value. 
So, for the selected phase space, the following values were obtained: $k_{T}^{\rm cut} = 5.8\, 
(6.8)\,{\rm GeV}$ 
for $^3P^{[1]}_1$ state and $k_{T}^{\rm cut} = 0.7\, (0.9)\,{\rm GeV}$ for
$^3P^{[1]}_2$ state at $\sqrt{s} = 7\,(13)$ TeV. 
The difference in the $k_{T}^{\rm cut}$ values for $^3P^{[1]}_1$ and $^3P^{[1]}_2$ states 
can be mainly attributed to the different behaviour of the corresponding production amplitudes 
at low transverse momenta.

It is important to note that the matched $2 \to 1$ and $2 \to 2$ cross sections very weakly 
depend on the exact $k_{T}^{\rm cut}$ values.
In fact, some reasonable variation in $k_{T}^{\rm cut}$ by $\pm\,0.5$ GeV around its central 
value (pink bands in Fig.~\ref{fig:kt_cut}) 
results in a negligible difference in the LO $+$ NLO$^*$ predictions, as it will be demonstrated below.
Thus, the uncertainties connected with the $k_{T}^{\rm cut}$ choice are rather small and
can be safely neglected in comparison with the ones coming, for example, from the standard 
scale variations.
 
\begin{figure}
\begin{center}
{\includegraphics[width=.5\textwidth]{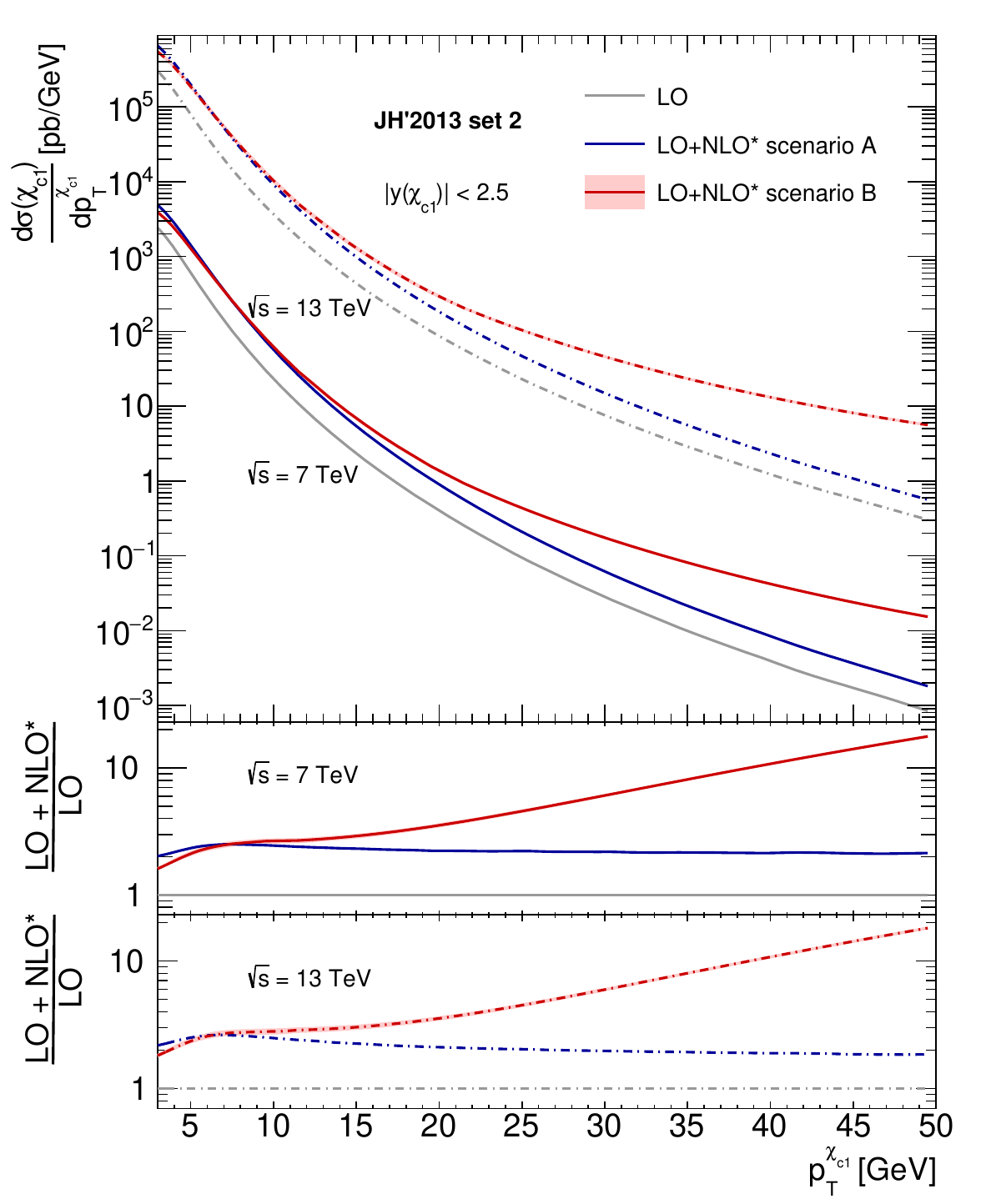}}\hfill
{\includegraphics[width=.5\textwidth]{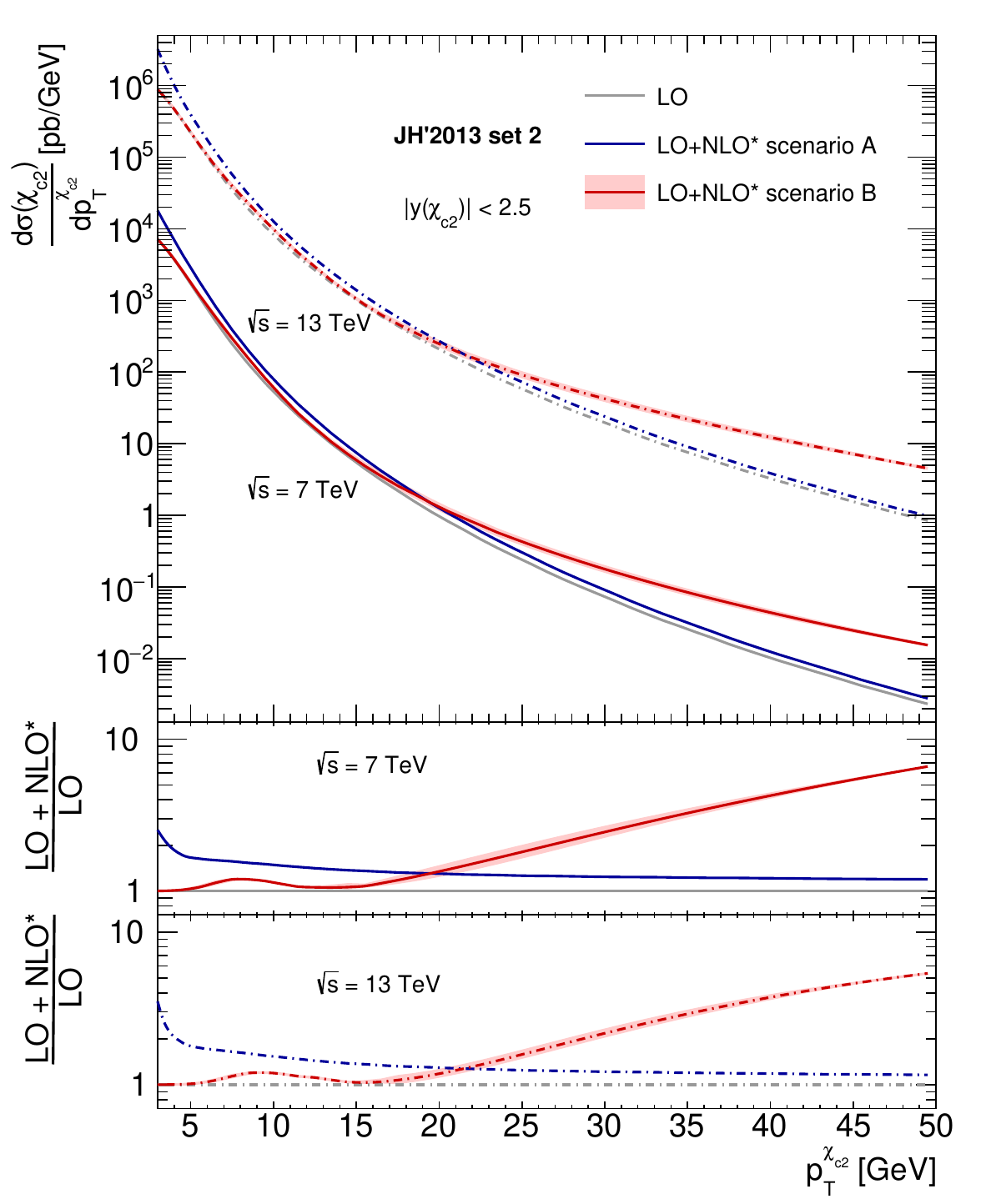}}\hfill
{\includegraphics[width=.5\textwidth]{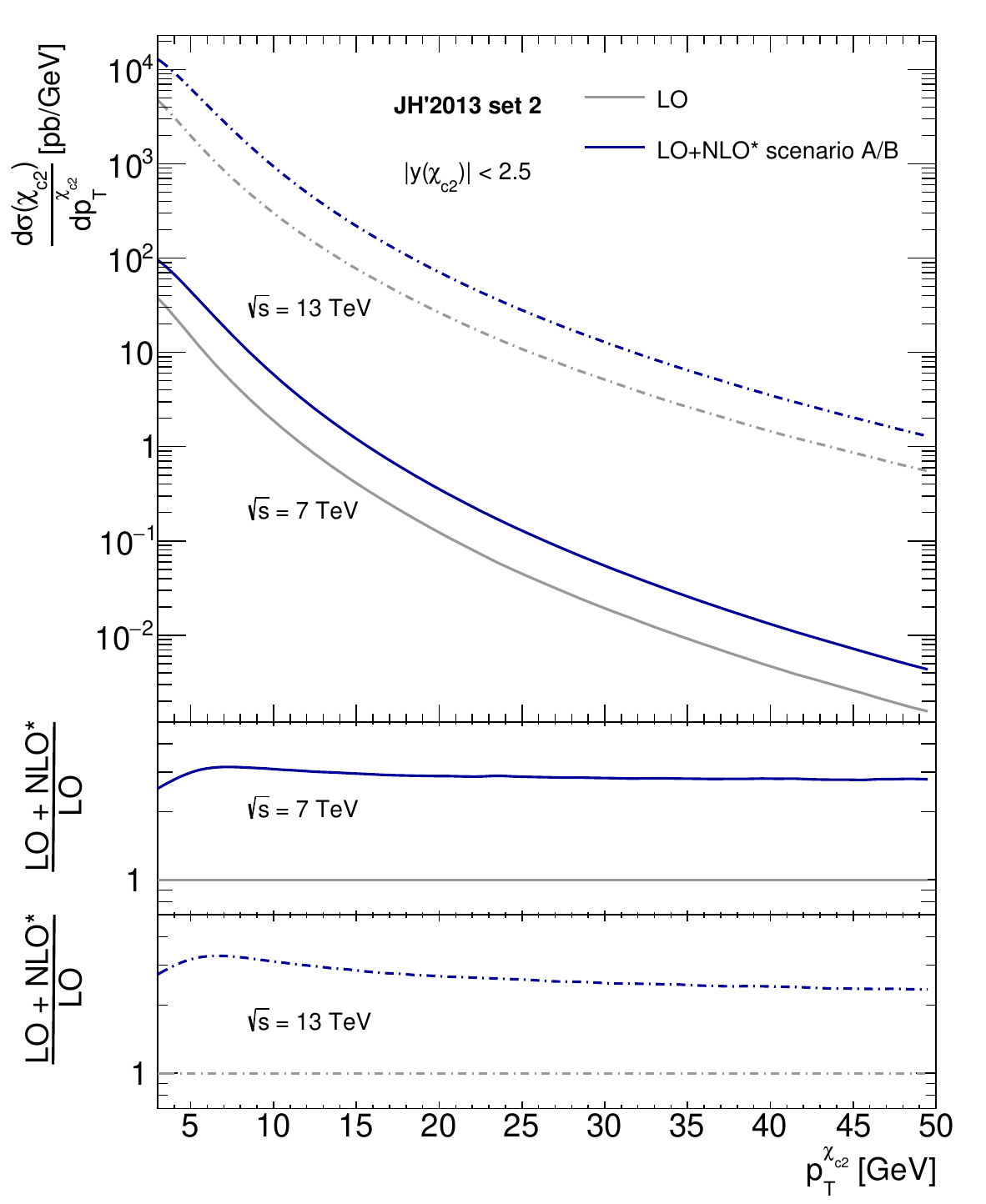}}\hfill
\caption{A comparison between the differential cross sections
for $\chi_{c1}$ mesons produced in the $^3P^{[1]}_{1}$ channel (left panel), 
$\chi_{c2}$ mesons produced in the $^3P^{[1]}_2$ (right panel) 
and $^3S^{[8]}_1$ states (lower panel)
in the central rapidity region $|y(\chi_{cJ})| < 2.5$ at $\sqrt{s} = 7$ and 
$13$~TeV ($\times 100$)
for merging scenarios A and B, and pure LO calculations. 
The shaded pink band represents the uncertainties coming from $k_{T}^{\rm cut}$
variations indicated in Fig.~\ref{fig:kt_cut}. The LDMEs are taken from Table~\ref{tab:LDMEs}.}
\label{fig:NLO_comparison}
 \end{center}
\end{figure}

Now we turn to a numerical comparison between the different merging scenarios and to a
comparison of the LO $+$ NLO$^*$ predictions with the pure LO calculations. 
The $^3P^{[1]}_{J}$ and $^3S^{[8]}_{1}$ contributions are separately shown in Fig.~\ref{fig:NLO_comparison} 
as functions of the produced $\chi_{cJ}$ meson transverse momentum for 
$\sqrt{s} = 7$ and $13$~TeV. 
We find that the difference between the merging scenarios becomes well pronounced
at large transverse momenta $p_T(\chi_{cJ})$, while both scenarios A and B
lead to close results for $^3P^{[1]}_{J}$ spectra
in the region of relatively low $p_{T}(\chi_{cJ}) < 20$~GeV.
The difference observed at high $p_T(\chi_{cJ})$ can probably be attributed to the role
of diagrams where gluons are emitted from the quark line.
In the scenario A, such diagrams are suppressed by the DCE cut, 
while they are taken into account in the scenario B.
The difference between the LO and LO $+$ NLO$^*$ predictions
for the $^3P^{[1]}_{2}$ spectra in the scenario A is rather small.

 
\section{Numerical results} \indent

In this section, we present the results of our calculations and perform a comparison with available Tevatron and LHC data.
In contrast with previous calculations\cite{Our-Charmonia-1, Our-Charmonia-3},
we preserve here the HQSS relations for the color singlet and color octet LDMEs:
\begin{gather}
\langle\mathcal{O}^{\chi_{cJ}}[^3P^{[1]}_{J}]\rangle\; =\; 6N_c(2J+1)\,\frac{|\mathcal{R}^{'\chi_{c0}}(0)|^2}{4\pi}, \nonumber \\
\langle\mathcal{O}^{\chi_{cJ}}[^3S^{[8]}_1]\rangle\; =\; (2J+1)\,\langle\mathcal{O}^{\chi_{c0}}[^3S^{[8]}_1]\rangle,
\label{eq:HQSS}
\end{gather}
\noindent
where $|\mathcal{R}^{'\chi_{c0}}(0)|^2 = 0.075$~GeV$^5$ is
the squared derivative of the $\chi_c$ color singlet wave function at the origin\cite{PotentialModelCalcutations-1, PotentialModelCalcutations-2}. 
The value of the color octet LDME, $\langle \mathcal{O}^{\chi_{c0}}[^3S^{[8]}_1]\rangle$, was extracted 
from a simultaneous best fit to the Tevatron and LHC data under requirement that it should be positive. 
This requirement follows from the approach\cite{TransitionMechanism} 
used to describe the transitions of a color octet $c\bar c$ pair into 
a final state meson.
We use the following data sets: ATLAS measurements of the $\chi_{c1}$ and $\chi_{c2}$ 
transverse momentum distributions at $\sqrt s = 7$~TeV\cite{chic-ATLAS} and
CDF data on the $\chi_{c1} + \chi_{c2}$ combined spectra measured at $\sqrt s = 1.8$~TeV
as functions of the $J/\psi$ transverse momentum after the radiative decay
$\chi_c\to J/\psi+\gamma$\cite{chic-CDF}. 
The results of our combined fit with corresponding ${\chi}^{2}/{\rm n.d.f.}$ 
for different TMD gluon densities in a proton
are collected in the Table~\ref{tab:LDMEs}.
The LDMEs derived through the merging scenarios A and B
differ from each other. However, for the A0 and LLM'2022 gluon distributions
they more or less coincide within the fit uncertainties.

\begin{table}[ht]
\centering
\caption{The fitted values of NME $\langle \mathcal{O}^{\chi_{c0}}[^3S^{[8]}_1]\rangle/{\rm GeV}^3$.}
\begin{tabular}[t]{lcccc}
\hline
&Scenario A&${\chi}^{2}/{\rm n.d.f.}$&Scenario B&${\chi}^{2}/{\rm n.d.f.}$\\
\hline
JH'2013 set 2&(3.1\,$\pm$\,0.9)$\,\times\, 10^{-4}$&0.78&(1.7\,$\pm$\,0.6)\,$\times\, 10^{-4}$&0.39\\
A0&(1.9\,$\pm$\,1.9)$\,\times\, 10^{-4}$&1.8&(1.3\,$\pm$\,0.5)$\,\times\, 10^{-4}$&0.65\\
LLM'2022&(4.8\,$\pm$\,0.9)$\,\times\, 10^{-4}$&1.09&(3.9\,$\pm$\,0.8)$\,\times\, 10^{-4}$&1.18\\
\hline
\end{tabular}
\label{tab:LDMEs}
\end{table}%

\begin{figure}
\begin{center}
{\includegraphics[width=.49\textwidth]{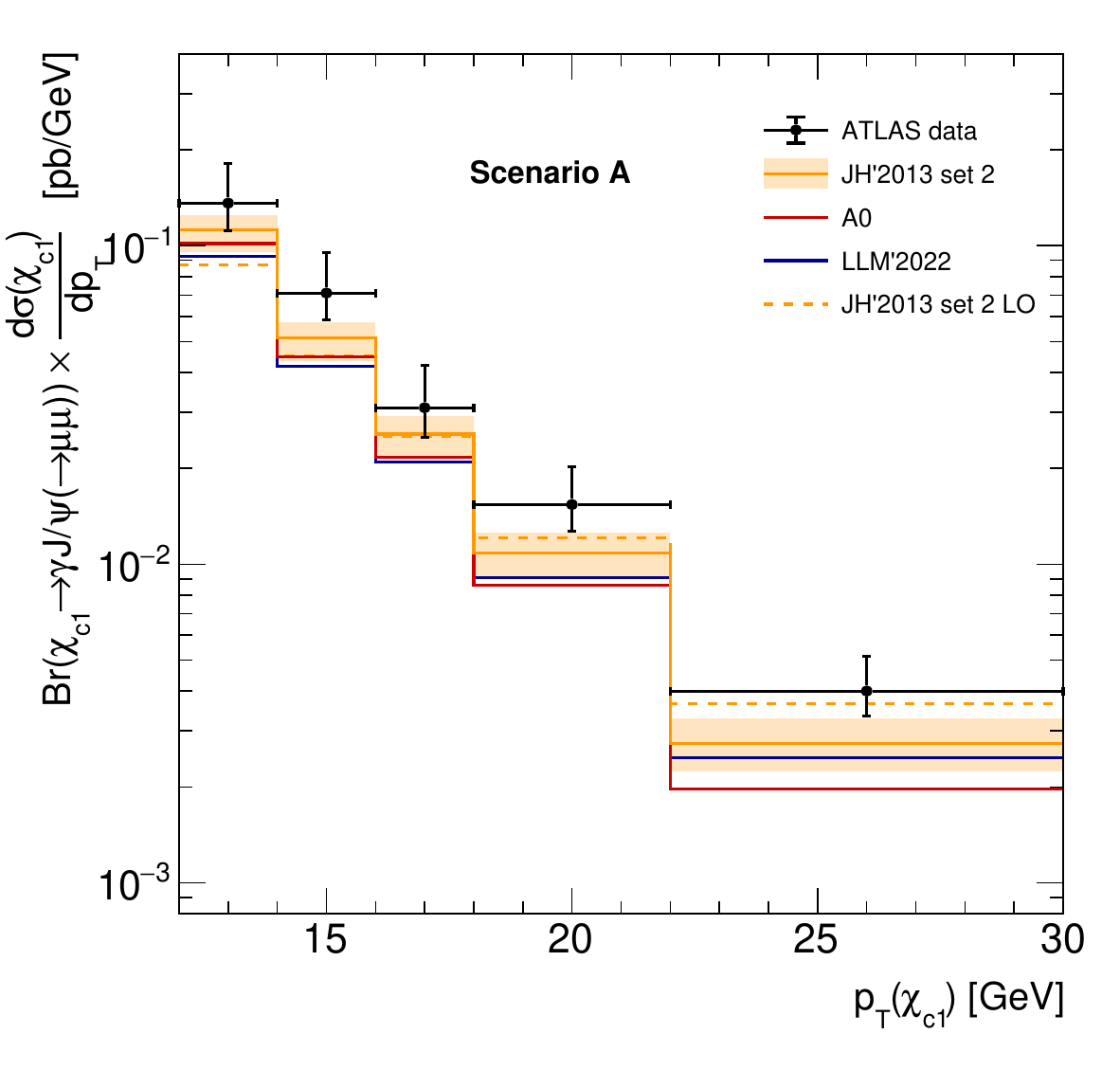}}\hfill
{\includegraphics[width=.49\textwidth]{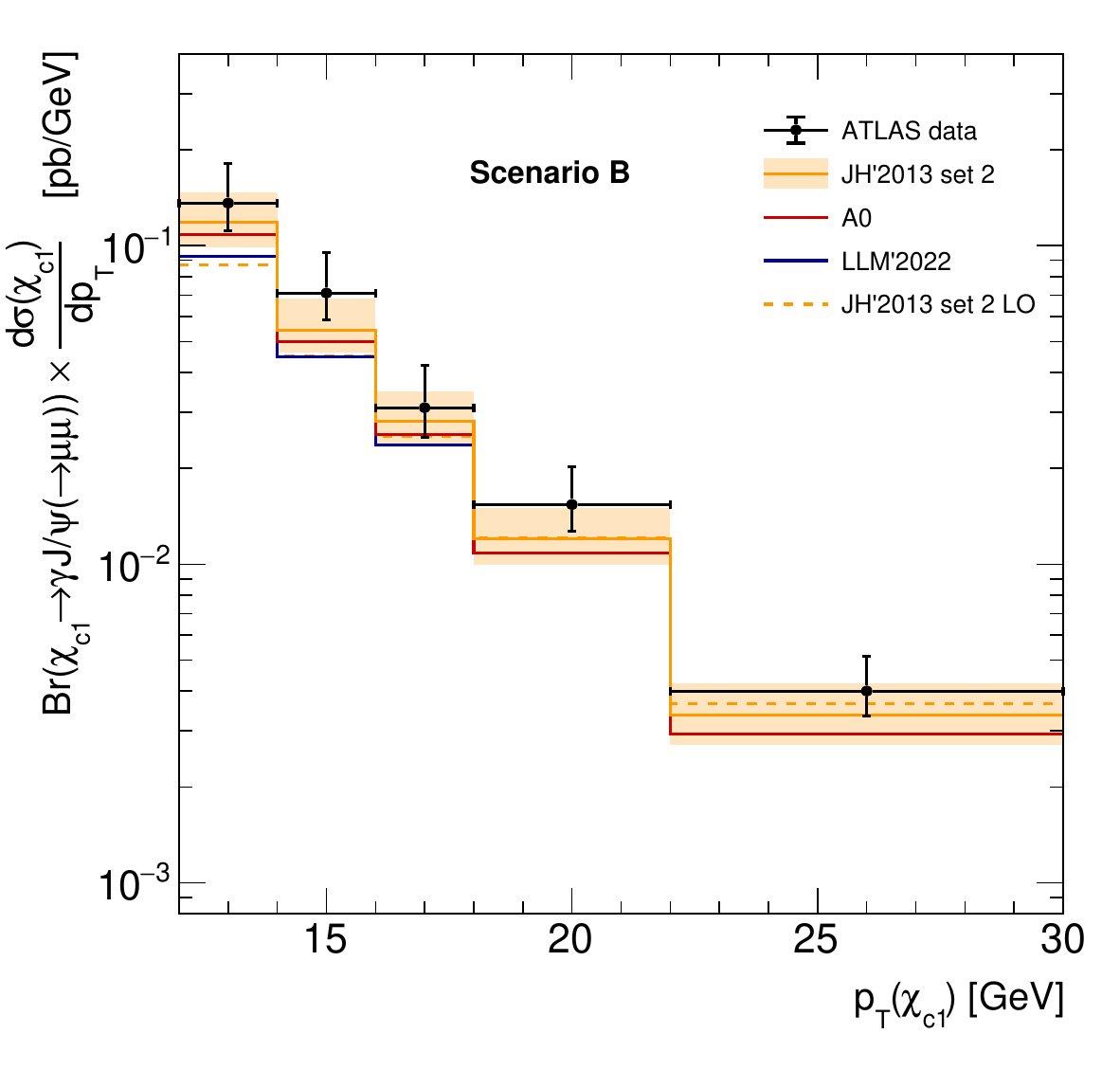}}\hfill
{\includegraphics[width=.49\textwidth]{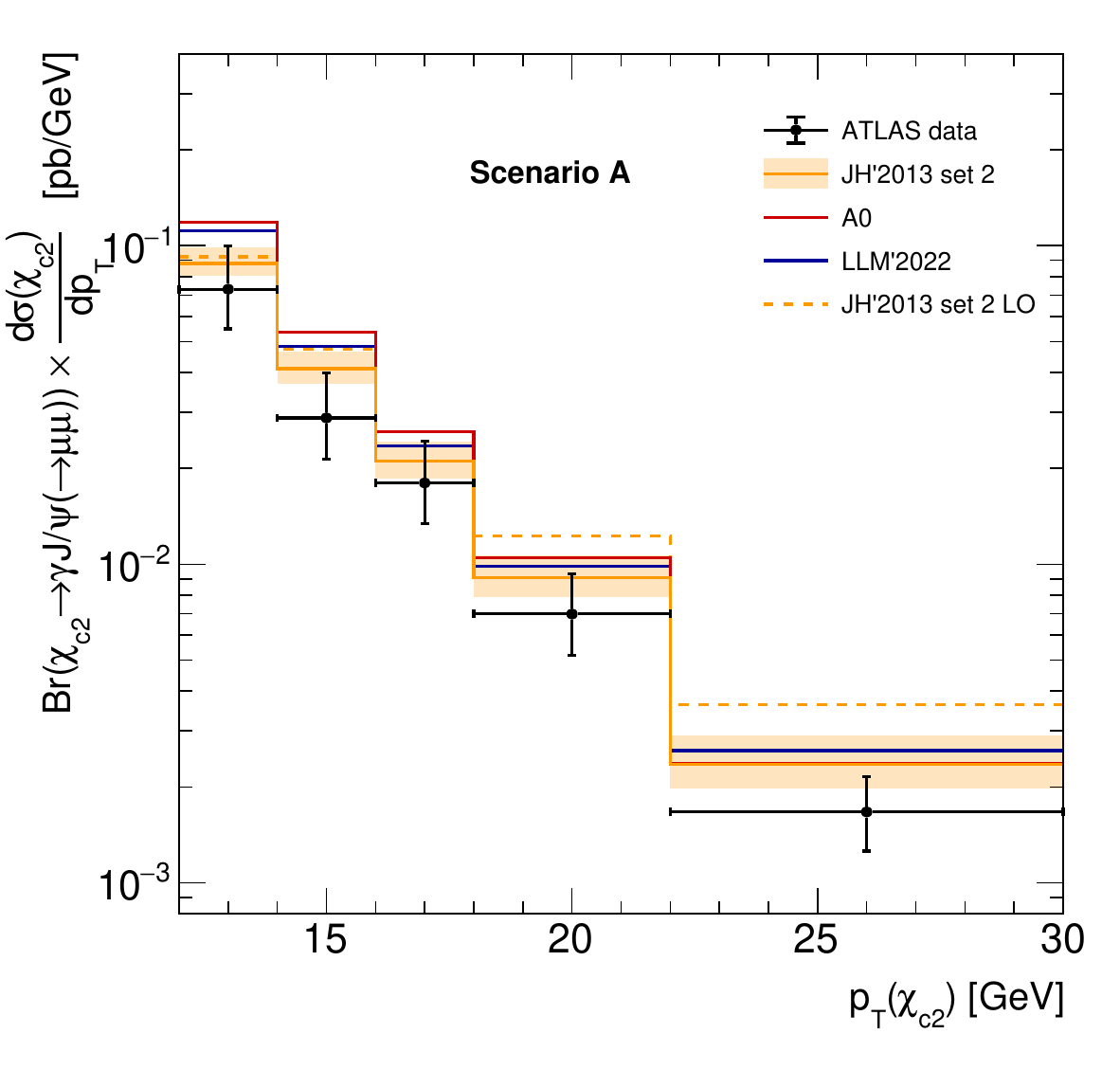}}\hfill
{\includegraphics[width=.49\textwidth]{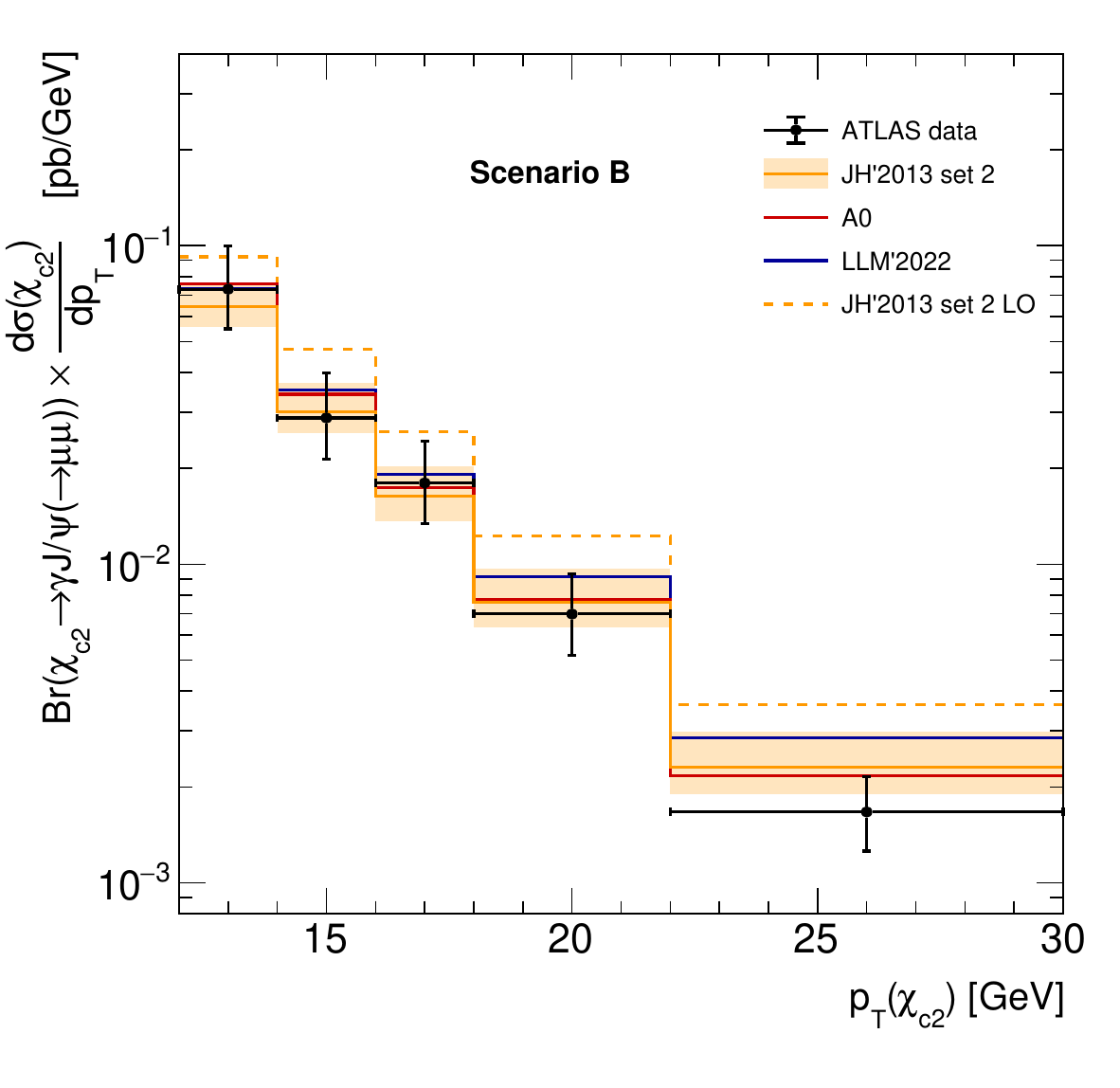}}\hfill
\caption{Differential cross sections for prompt $\chi_{c1}$ (upper panels) and $\chi_{c2}$ (lower panels) 
production in $pp$ collisions at $\sqrt{s} = 7$ TeV as functions of the $\chi_{c}$ 
transverse momentum. The kinematic cuts are described in the text. The ATLAS data are taken from\cite{chic-ATLAS}.}
\label{fig:ATLAS_chic}
 \end{center}
\end{figure}

\begin{figure}
\begin{center}
{\includegraphics[width=.49\textwidth]{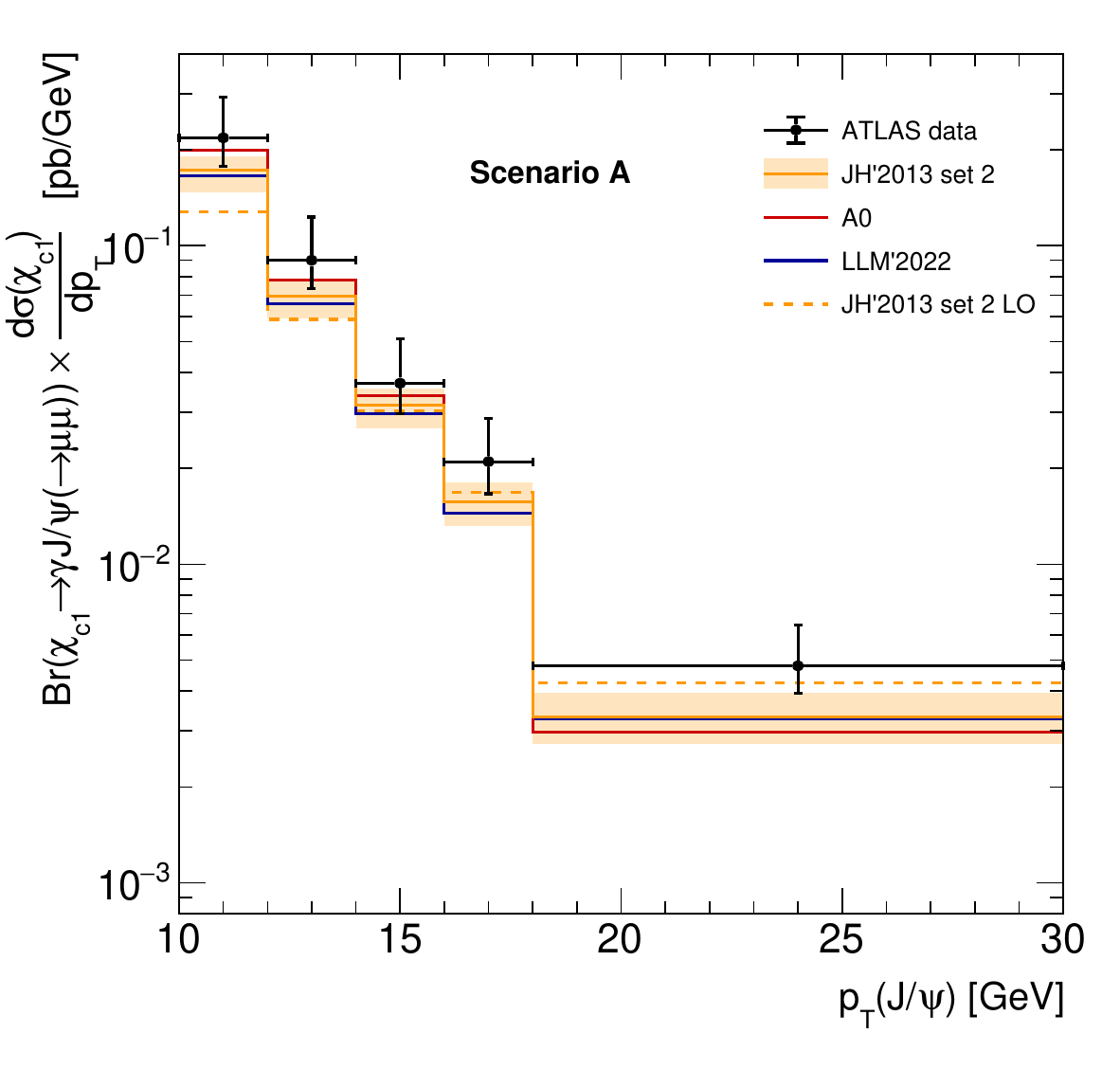}}\hfill
{\includegraphics[width=.49\textwidth]{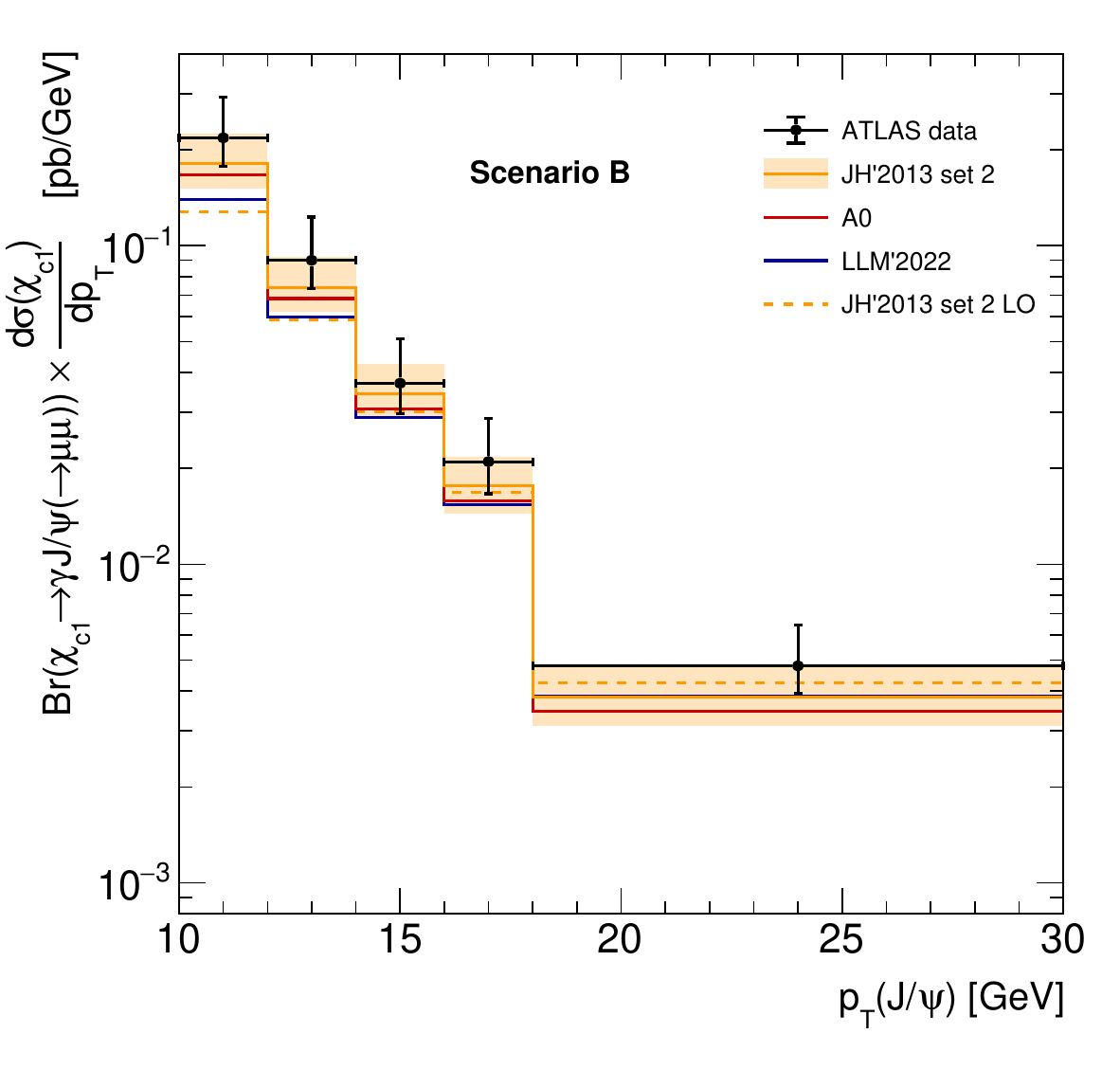}}\hfill
{\includegraphics[width=.49\textwidth]{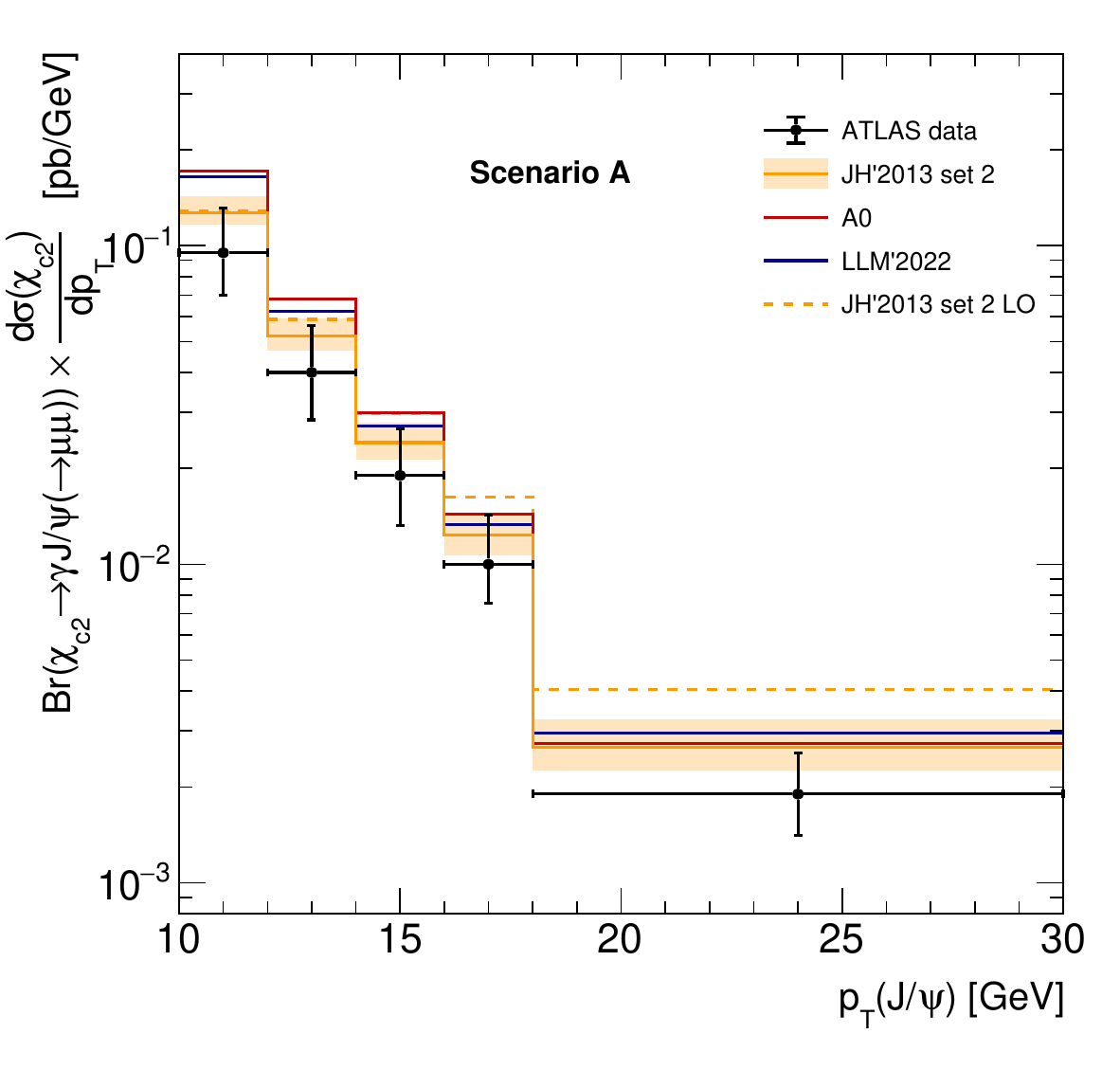}}\hfill
{\includegraphics[width=.49\textwidth]{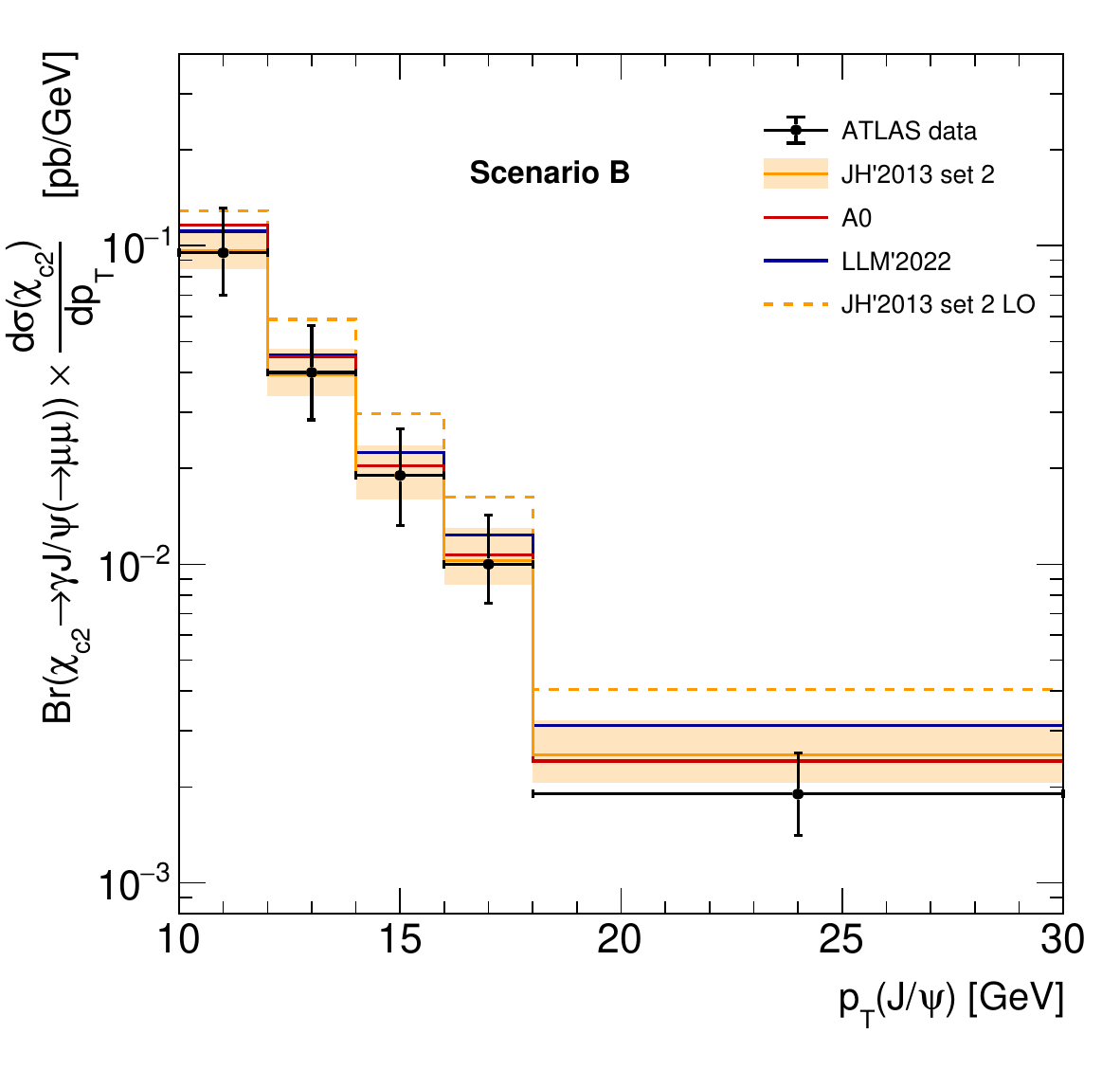}}\hfill
\caption{Differential cross sections for prompt $\chi_{c1}$ (upper panels) and $\chi_{c2}$ (lower panels) production in $pp$ 
collisions at $\sqrt{s} = 7$ TeV as functions of the decay $J/\psi$ transverse momentum.
The kinematic cuts are described in the text. The ATLAS data are taken from\cite{chic-ATLAS}.}
\label{fig:ATLAS_jpsi}
 \end{center}
\end{figure}

\begin{figure}
\begin{center}
{\includegraphics[width=.49\textwidth]{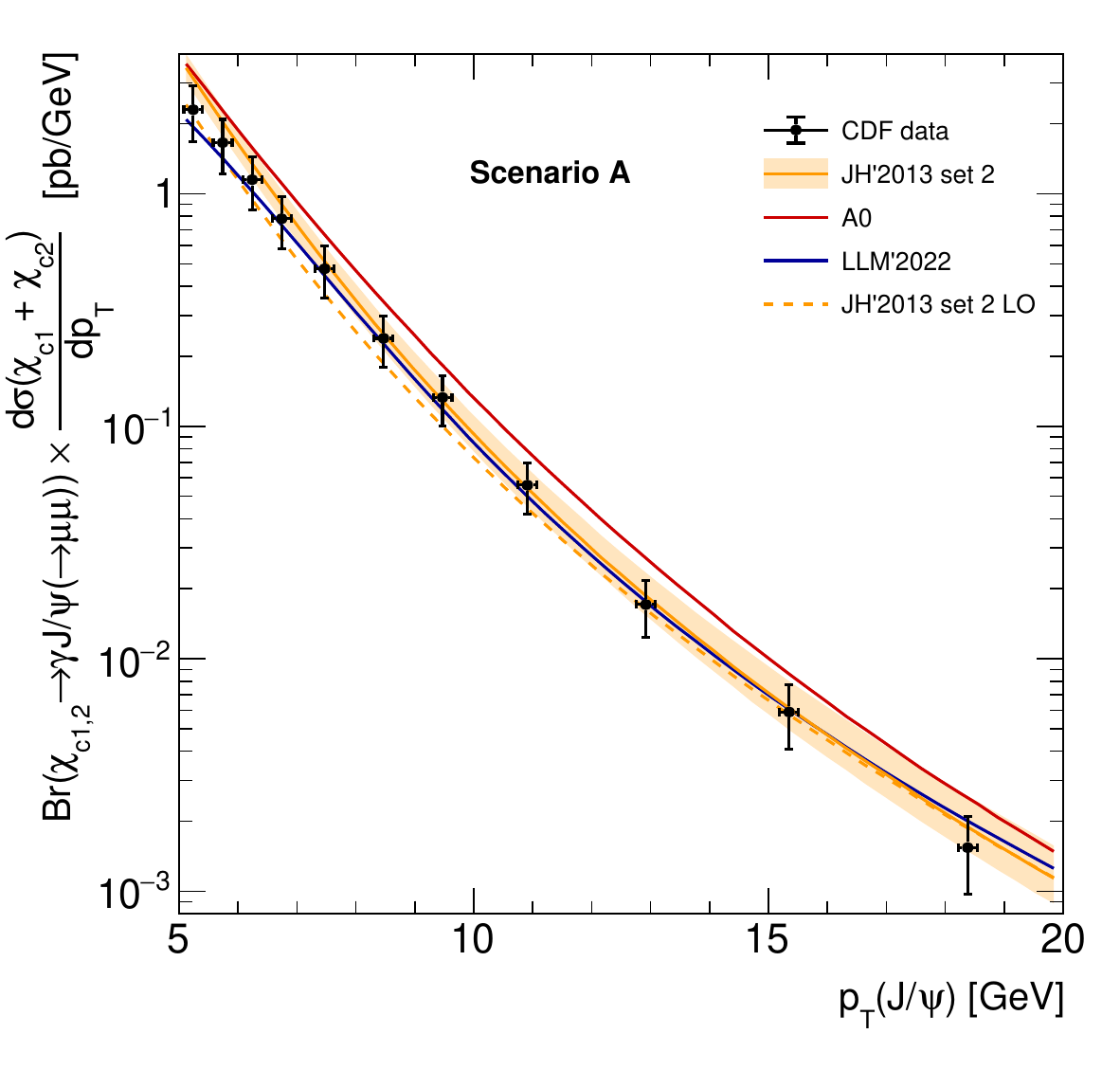}}\hfill
{\includegraphics[width=.49\textwidth]{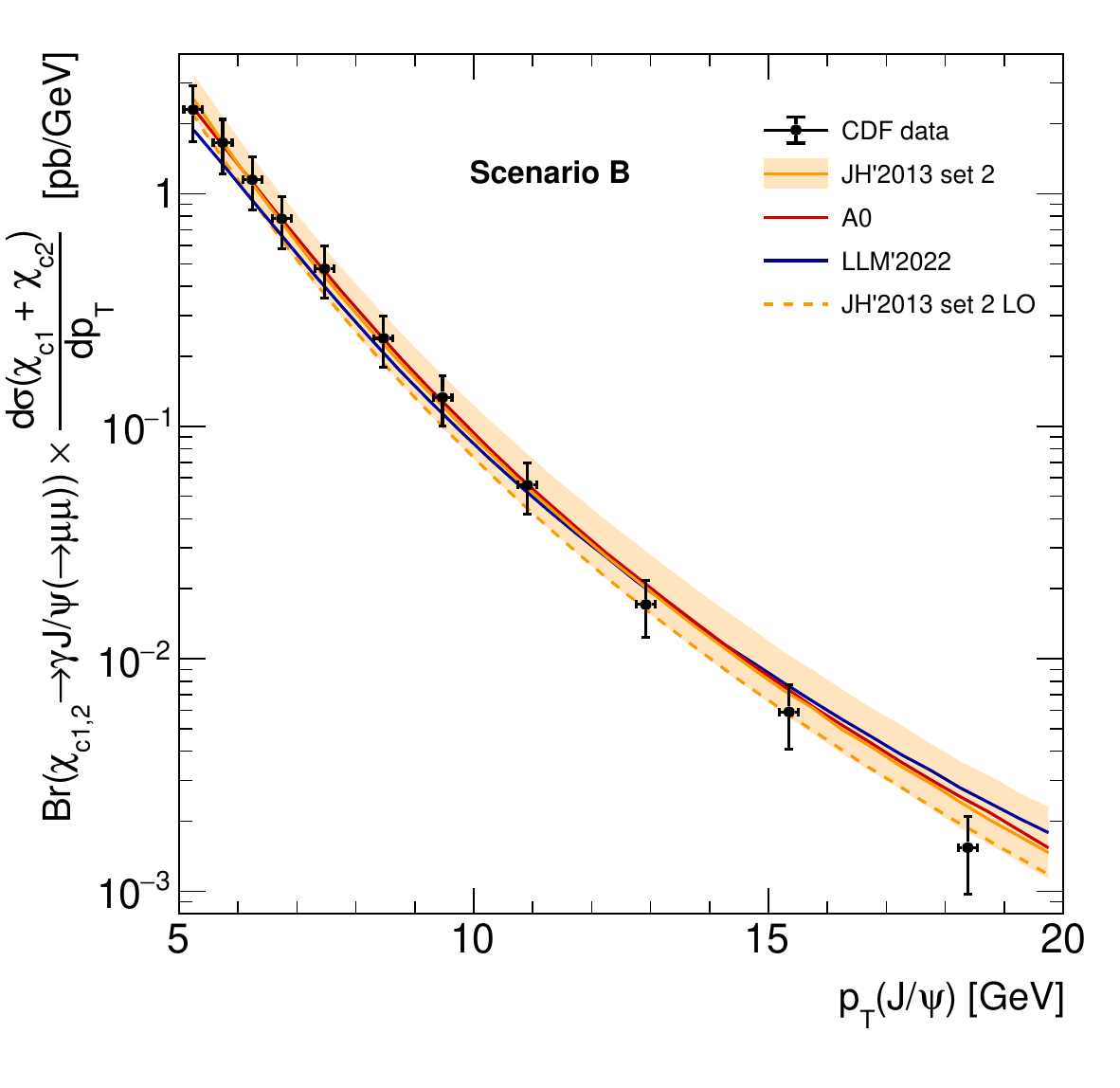}}\hfill
\caption{Differential cross section of prompt $\chi_{c}$ production in $p\bar{p}$ 
collisions at $\sqrt{s} = 1.8$ TeV as function of the decay $J/\psi$ transverse momentum.
The kinematic cuts are described in the text. The CDF data are taken from\cite{chic-CDF}.}
\label{fig:CDF_jpsi}
 \end{center}
\end{figure}

\begin{figure}
\begin{center}
{\includegraphics[width=.49\textwidth]{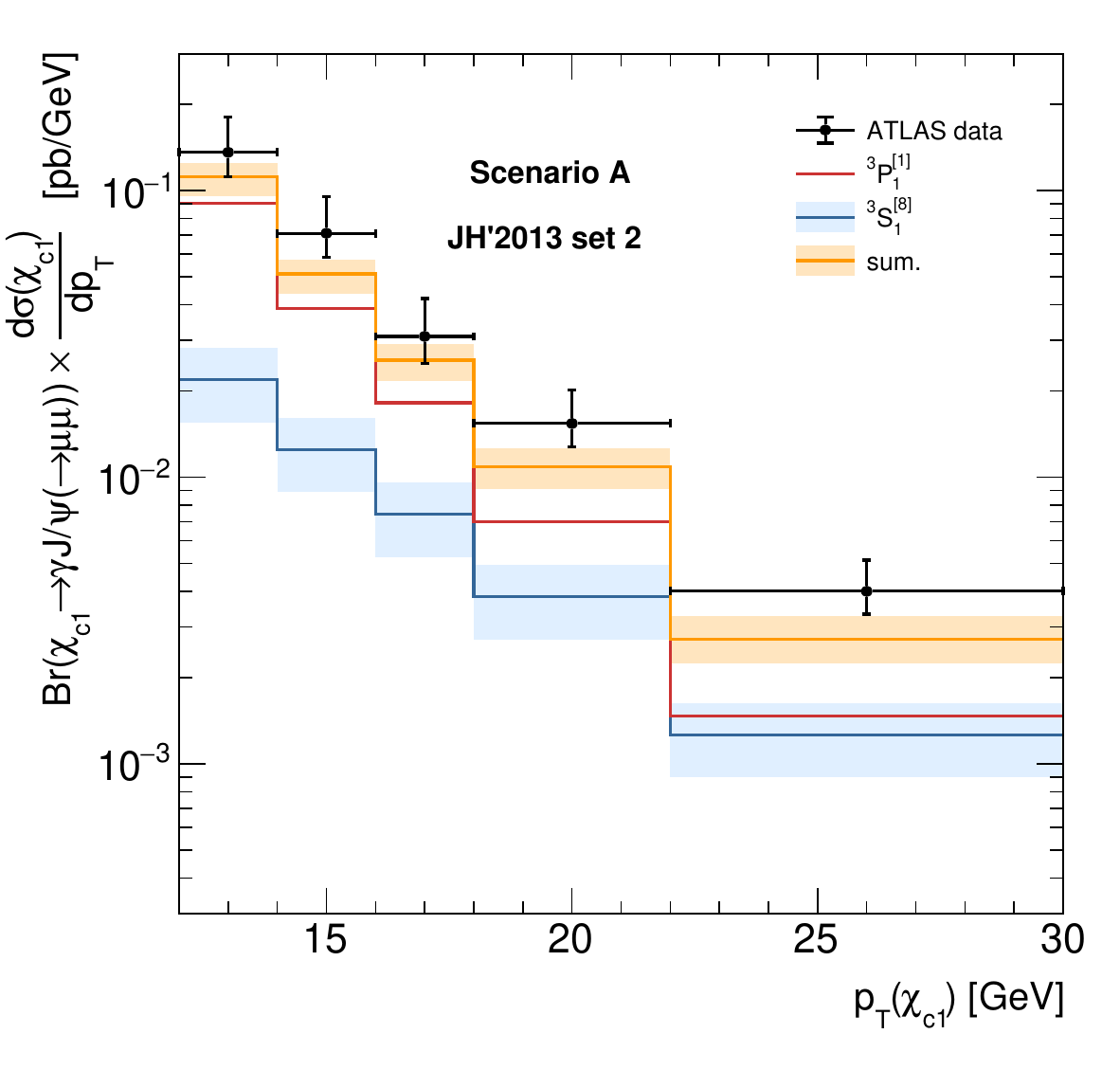}}\hfill
{\includegraphics[width=.49\textwidth]{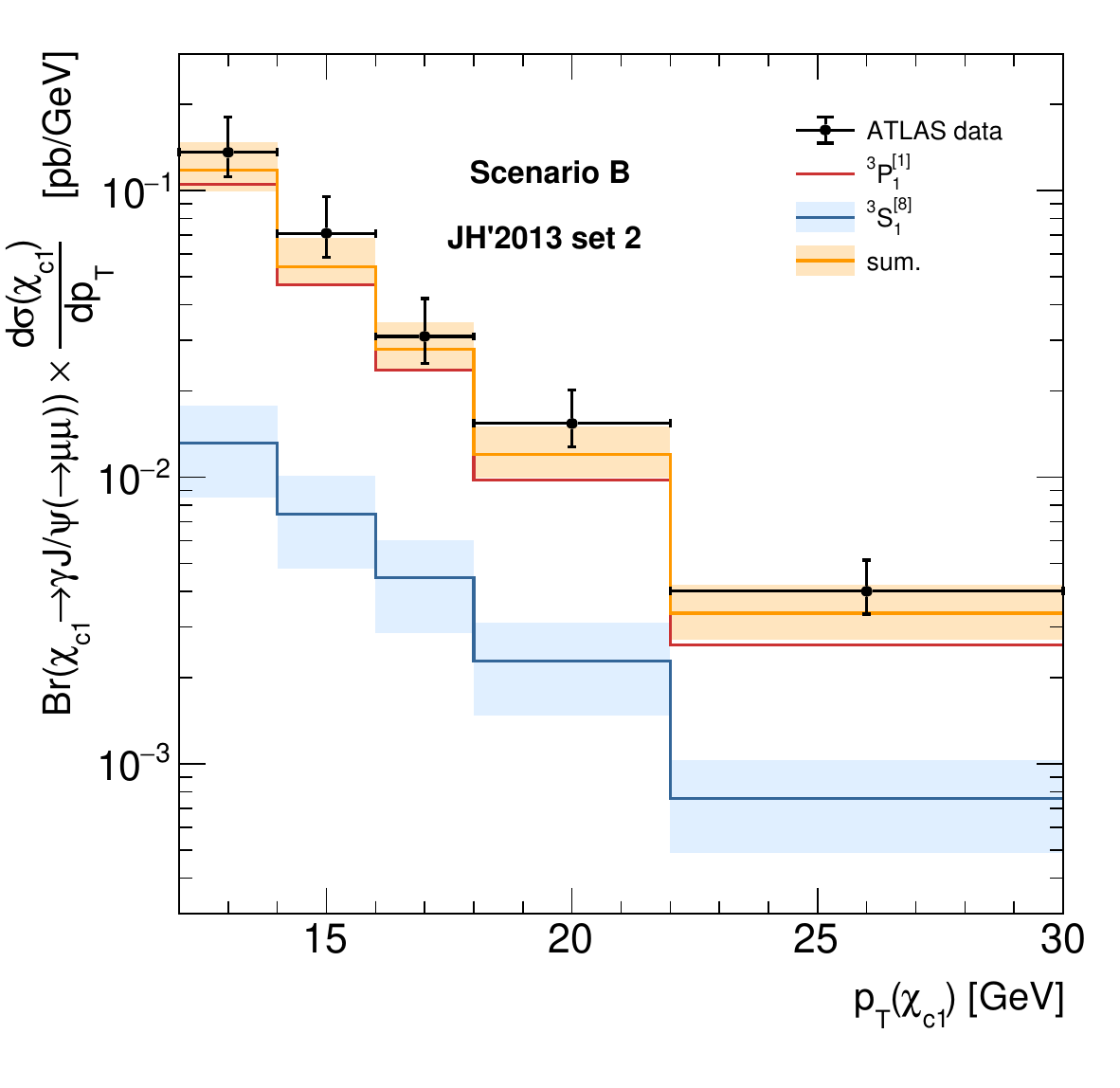}}\hfill
{\includegraphics[width=.49\textwidth]{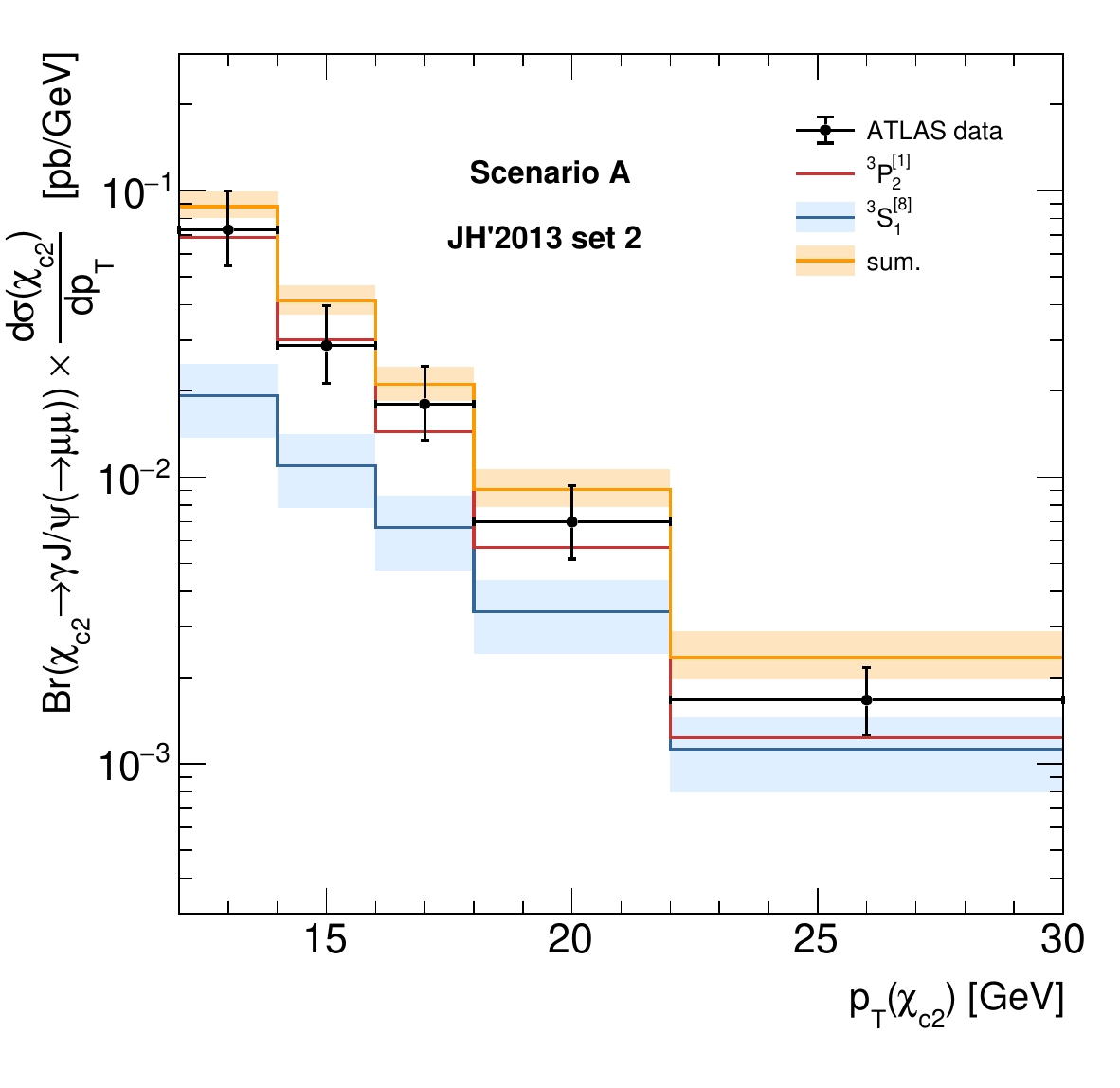}}\hfill
{\includegraphics[width=.49\textwidth]{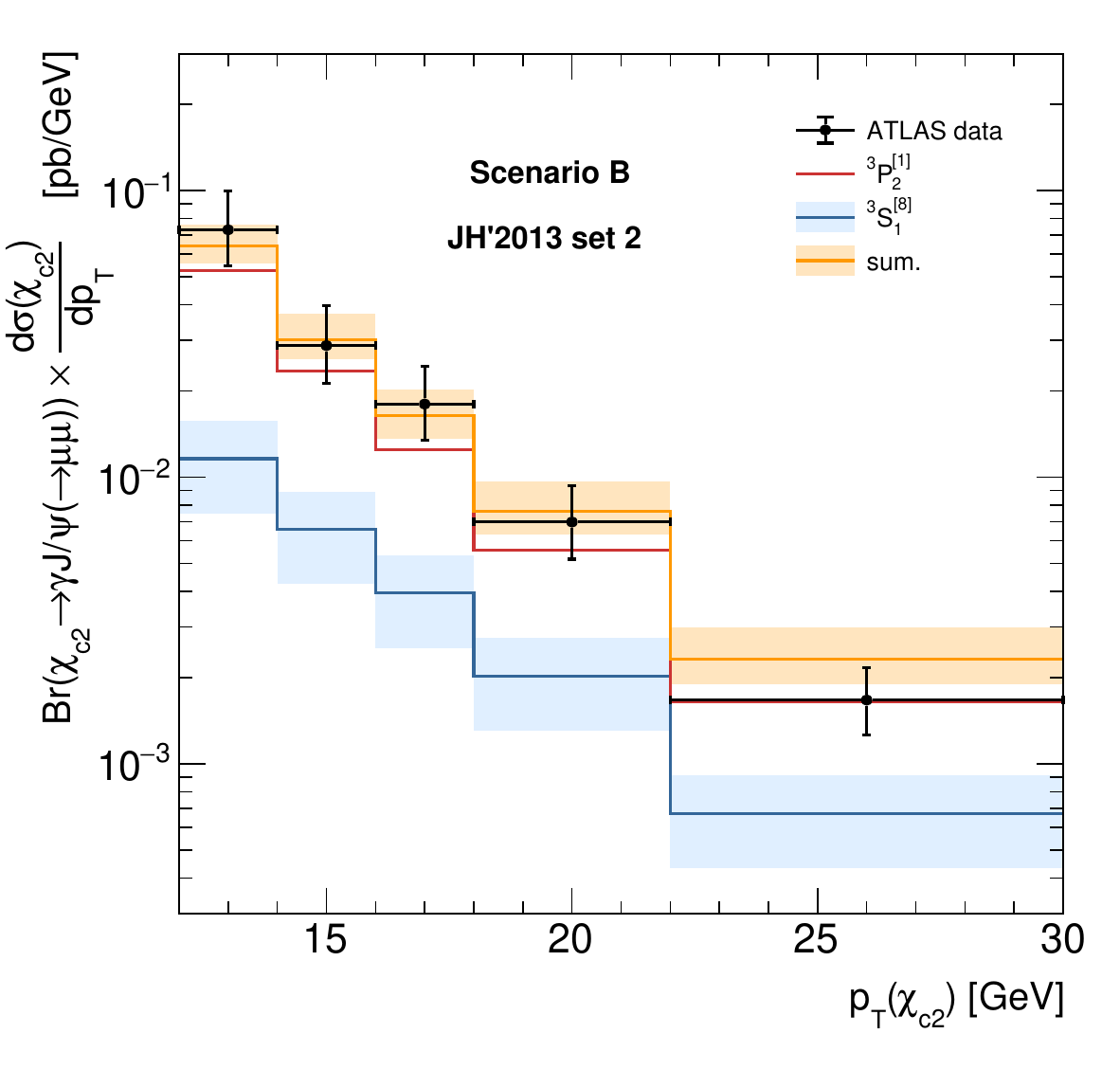}}\hfill
\caption{Different contributions to the $\chi_{c1}$ (upper panels) and $\chi_{c2}$ (lower panels) production cross sections in $pp$ 
collisions at $\sqrt{s} = 7$ TeV. The JH'2013 set 2 gluon density is used.
The kinematic cuts are described in the text. Blue shaded bands represent the uncertainties in the $\langle\mathcal{O} ^3S^{[8]}_1 \rangle$ 
determination (see Table~\ref{tab:LDMEs}). The ATLAS data are taken from\cite{chic-ATLAS}.}
\label{fig:ATLAS_details}
 \end{center}
\end{figure}

\begin{figure}
\begin{center}
{\includegraphics[width=.49\textwidth]{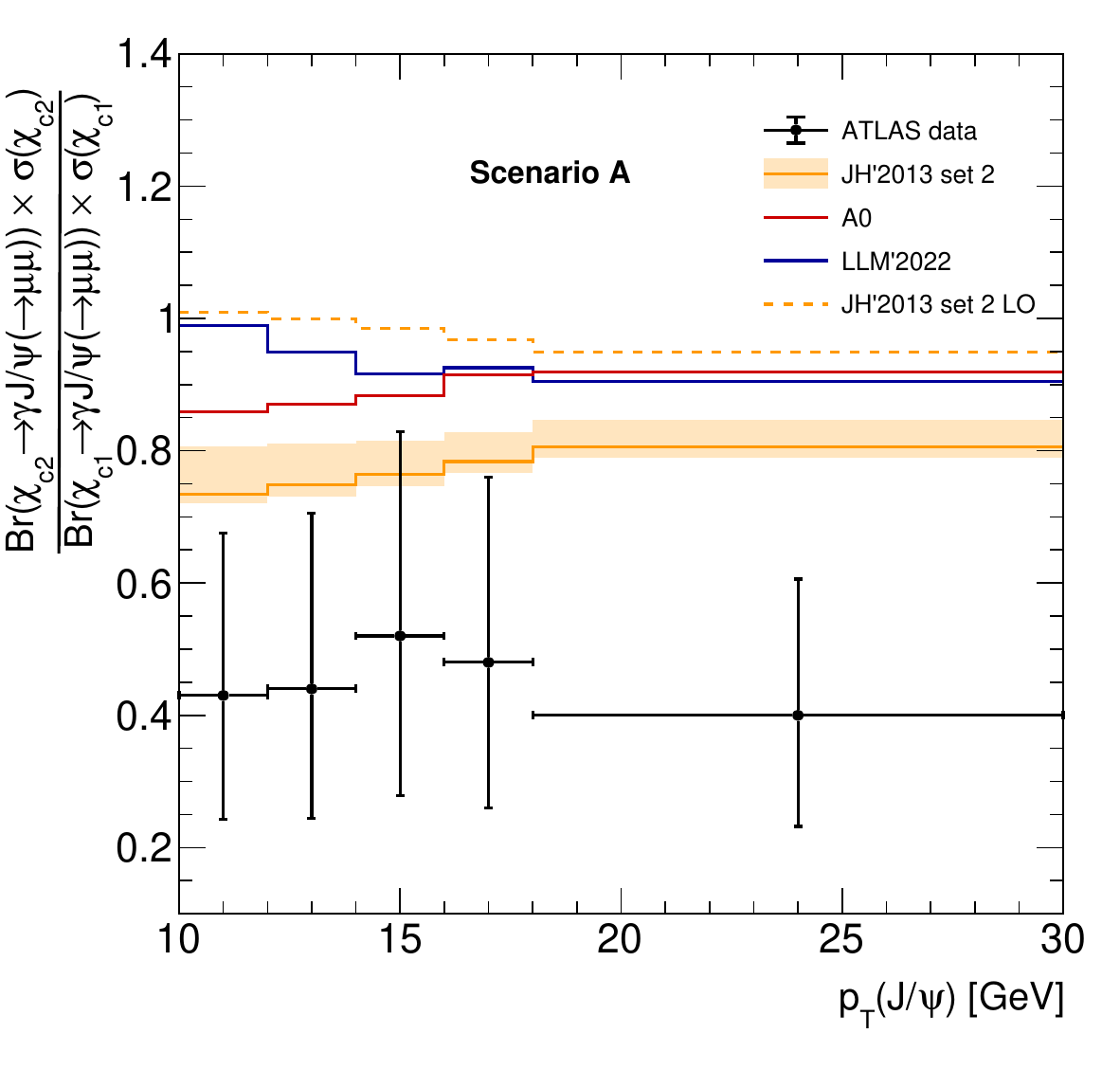}}\hfill
{\includegraphics[width=.49\textwidth]{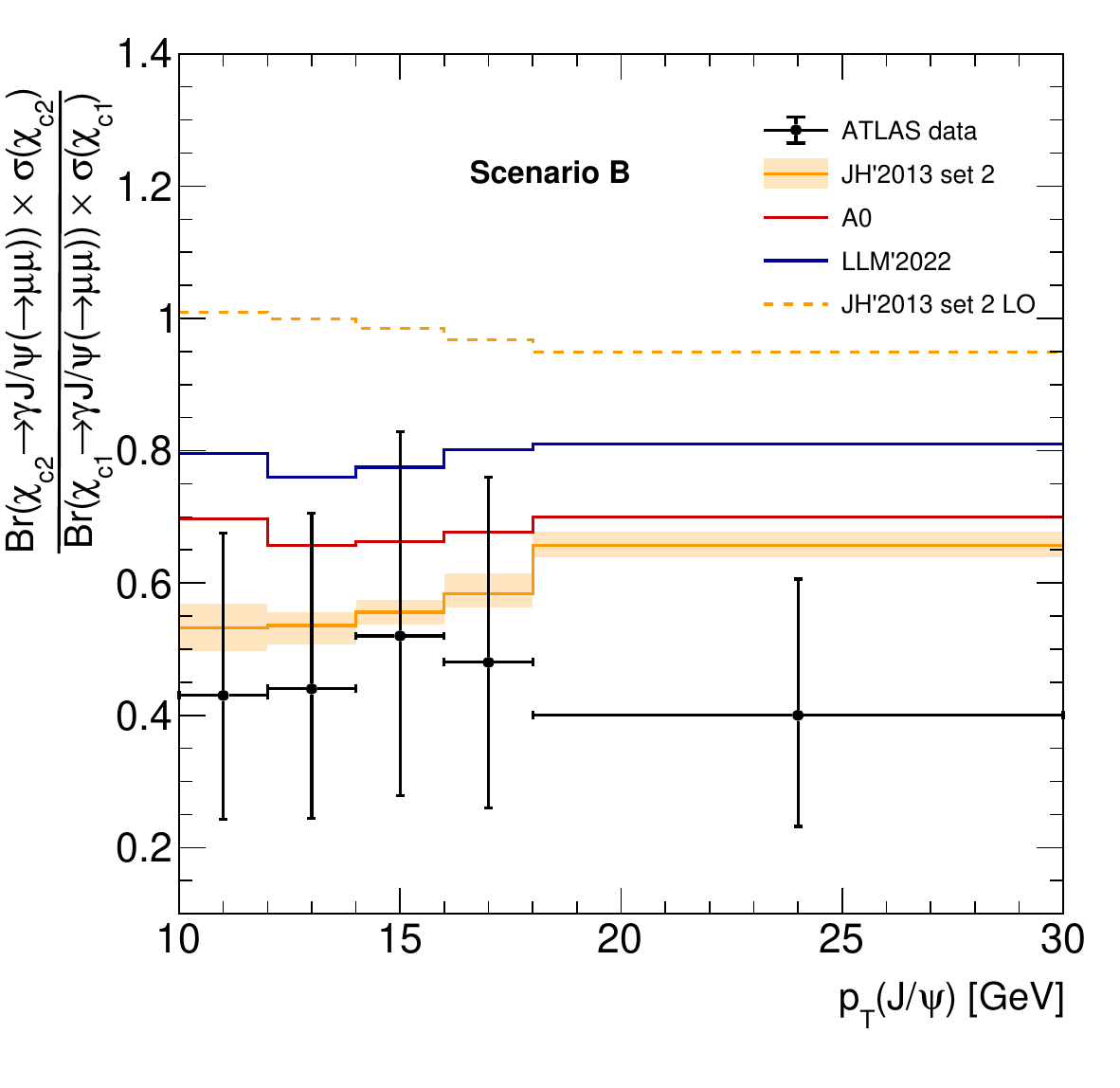}}\hfill
{\includegraphics[width=.49\textwidth]{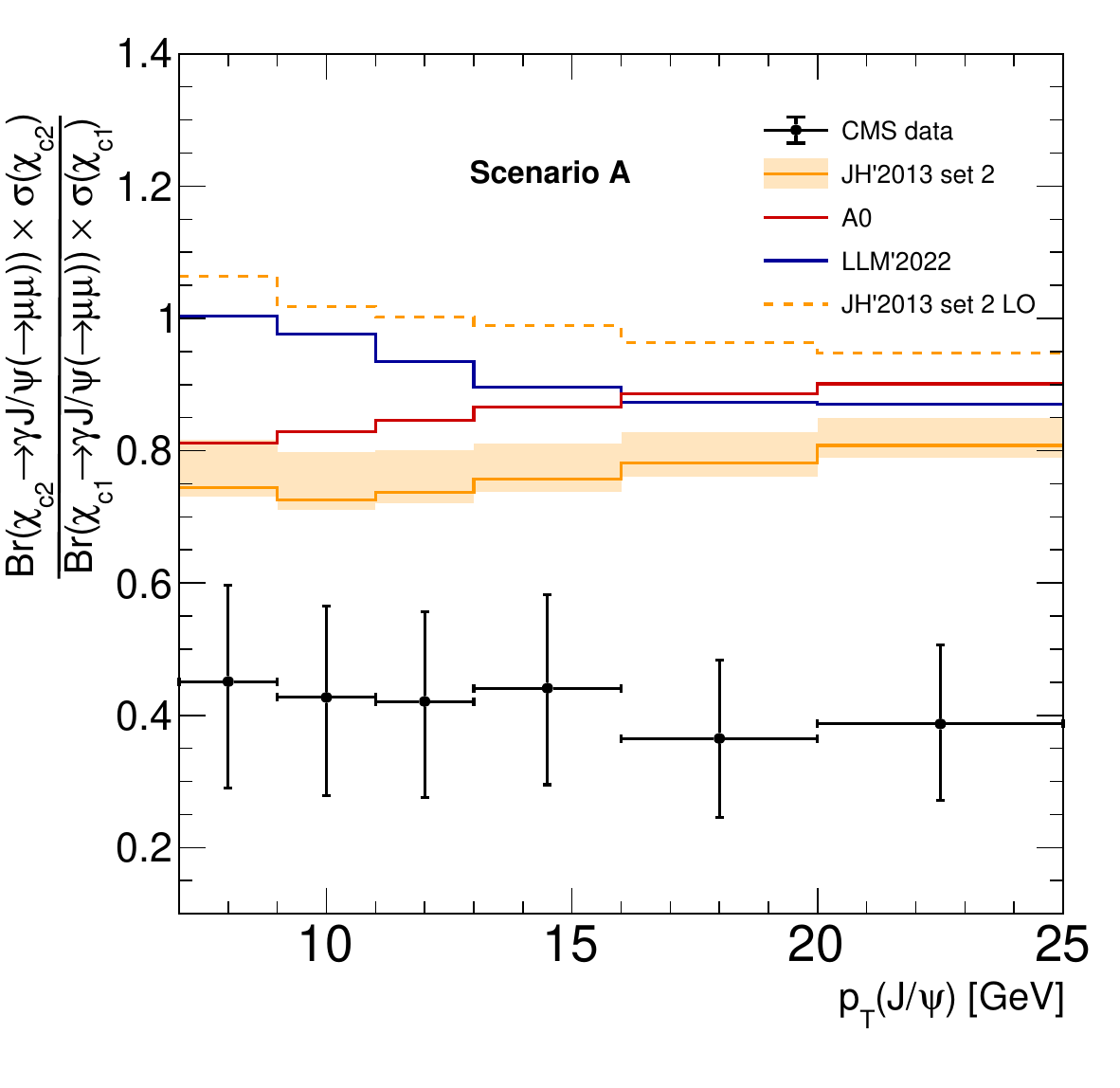}}\hfill
{\includegraphics[width=.49\textwidth]{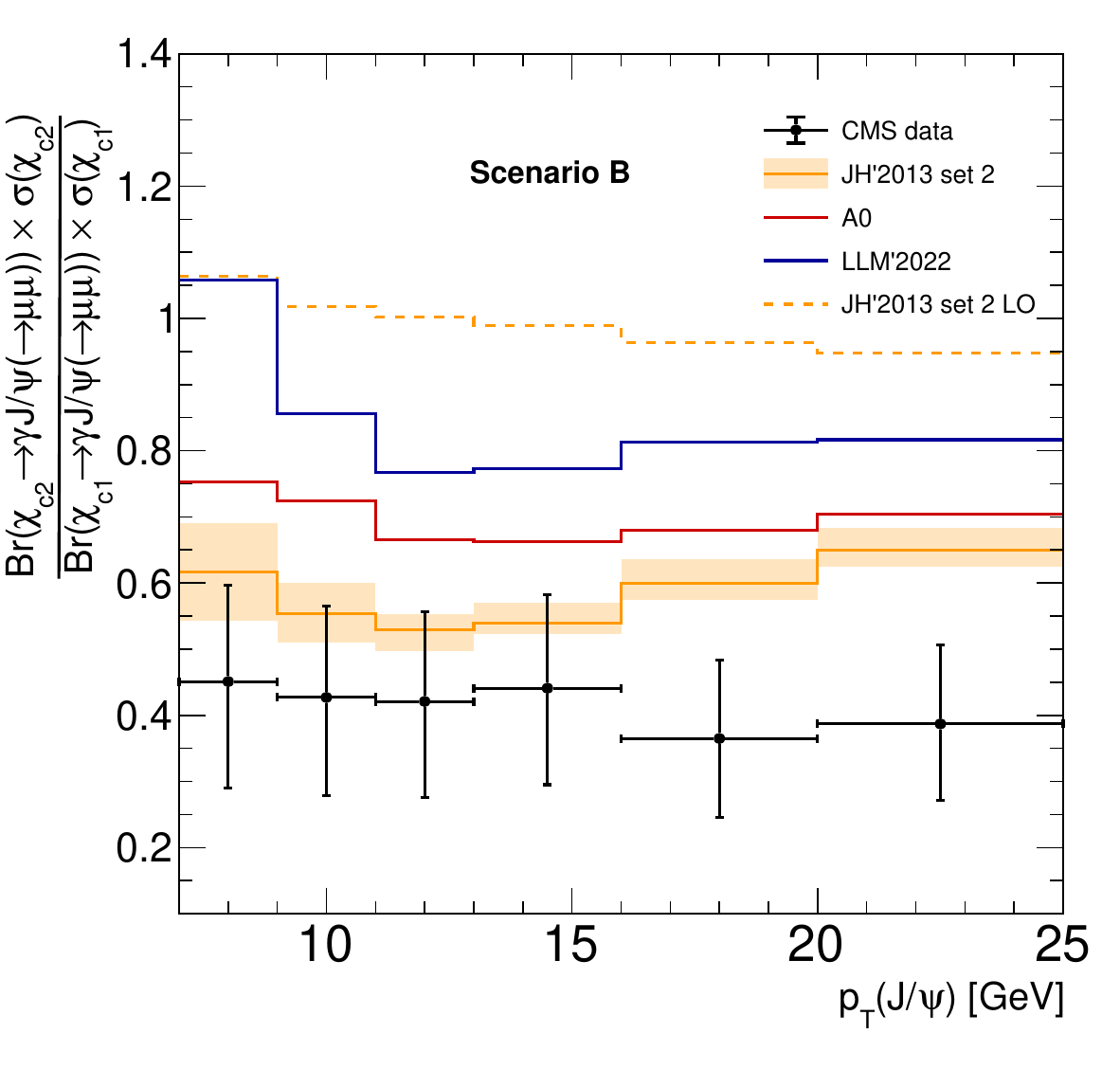}}\hfill
\caption{The ratio of the production rates $\sigma(\chi_{c2})/\sigma(\chi_{c1})$ calculated at 
$\sqrt{s} = 7$ TeV as function of the decay $J/\psi$ transverse momentum. 
The ATLAS and CMS data are taken from\cite{chic-ATLAS, chic-CMS-ratio}.}
\label{fig:ATLAS_ratio}
 \end{center}
\end{figure}

The transverse momentum distributions of $\chi_c$ mesons
obtained with the fitted LDMEs are shown in Figs.~\ref{fig:ATLAS_chic} --- \ref{fig:CDF_jpsi}.
Different contributions to the calculated cross sections
from the considered production mechanisms are separately shown in Fig.~\ref{fig:ATLAS_details}.
The shaded orange bands represent the estimated theoretical uncertainties for JH'2013 set 2 gluon density.
The latter contain the uncertainties in the $\langle\mathcal{O} ^3S^{[8]}_1 \rangle$ determination (see Table~\ref{tab:LDMEs}) and scale 
uncertainties estimated by varying the renormalization scale within a factor of two,
$\mu_R \to 2\mu_R$ and $\mu_R \to \mu_R/2$ (see\cite{JH2013} for more details). 
One can see that our predictions are in a reasonably good agreement with 
measured $\chi_c$ spectra within the theoretical and experimental uncertainties. 
However, we find that scenario B provides somewhat better description
of the ATLAS data in comparison with scenario A,
where contributions from the $^3S_1^{[8]}$ terms
play more important role (see Fig.~\ref{fig:ATLAS_details}).
In fact, scenario A leads to some underestimation of the measured $\chi_{c1}$
spectra and slight overestimation of the $\chi_{c2}$ data.
This immediately results in an overestimation of the $\sigma(\chi_{c2})/\sigma(\chi_{c1})$ production rates, 
whereas the predictions of scenario B are more close to the data, as it is demonstrated in Fig.~\ref{fig:ATLAS_ratio}.
The $\sigma(\chi_{c2})/\sigma(\chi_{c1})$ ratio is found to be sensitive to the TMD gluon
 density in a proton, and a reasonable description is achieved with JH'2013 set 2 gluon.
The CDF measurements are described well by both merging scenarios,
although the old A0 gluon distribution tends to overestimate the data in scenario A.

Additionally, in Figs.~\ref{fig:ATLAS_chic} --- \ref{fig:ATLAS_ratio} we show the results provided by the pure $2 \to 1$ calculations
where the HQSS relations~(\ref{eq:HQSS}) are taken into account\footnote{This is in contrast with the calculations\cite{Our-Charmonia-1, Our-Charmonia-3}.}.
One can see that such calculations lead to an unsatisfactory description of data\footnote{The $\chi^2/{\rm n.d.f.} \sim 2$ for JH'2013 set 2 
with fitted value $\langle\mathcal{O}^{\chi_{c0}}[^3S^{[8]}_1]\rangle = 1.94\times10^{-3}$ GeV$^3$.}. 
In particular, the incorrect shapes of the $p_T$ distributions
and noticeably overestimated $\chi_c$ relative production rates are observed, see Fig.~\ref{fig:ATLAS_ratio}.
The inclusion of NLO$^*$ terms significantly improves the overall agreement with the data in both scenarios and also leads to a good description of the measured 
$\sigma(\chi_{c2})/\sigma(\chi_{c1})$ ratio in scenario B.
Therefore, the previously stated violation of the HQSS relations for $\chi_c$ mesons \cite{Our-Charmonia-1, Our-Charmonia-3, UnEqualWaveFunctions}
can be explained by the absence of higher-order corrections in the ${\cal O}(\alpha_s^2)$ off-shell amplitudes.

\begin{figure}
\begin{center}
{\includegraphics[width=.49\textwidth]{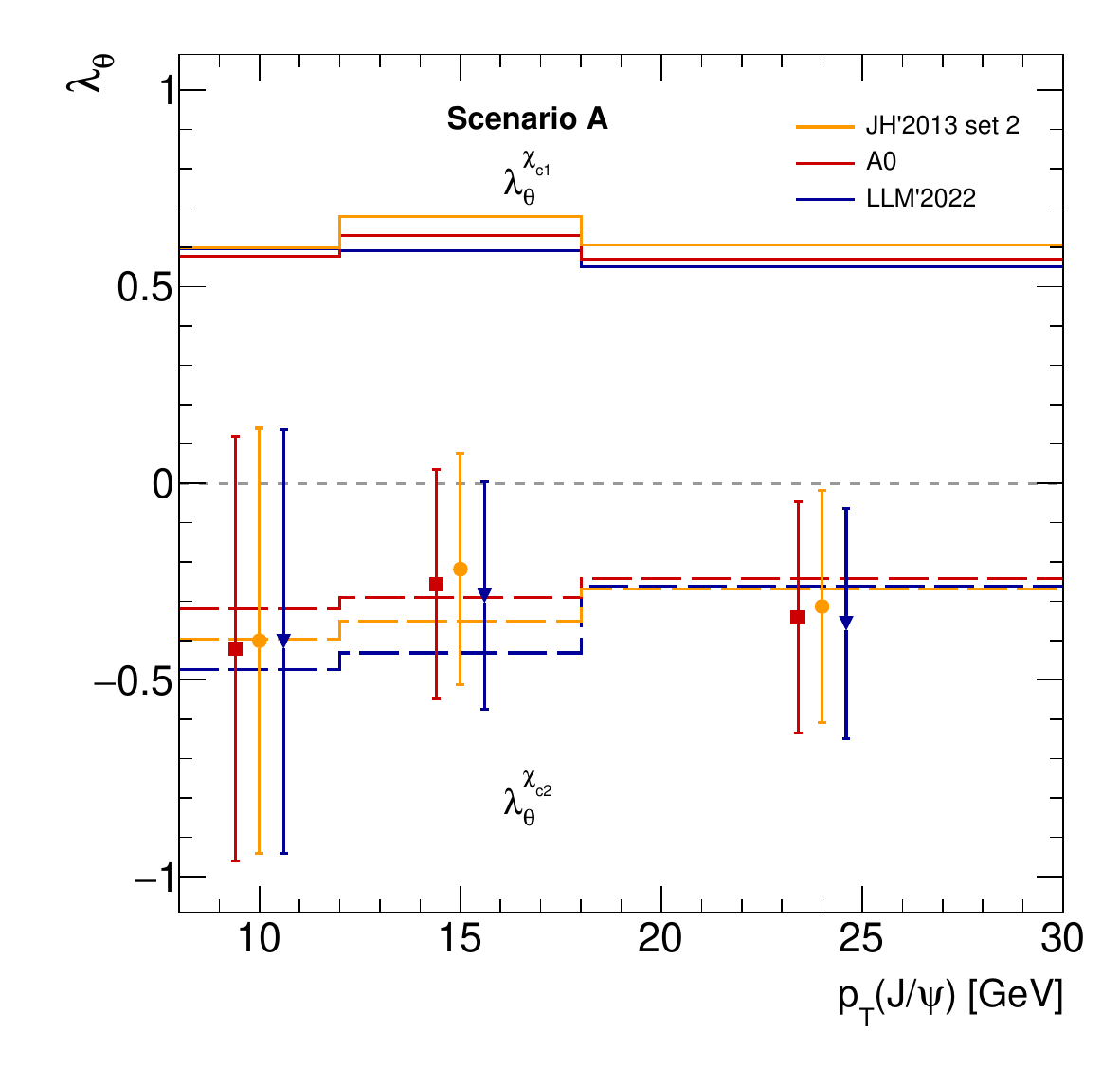}}\hfill
{\includegraphics[width=.49\textwidth]{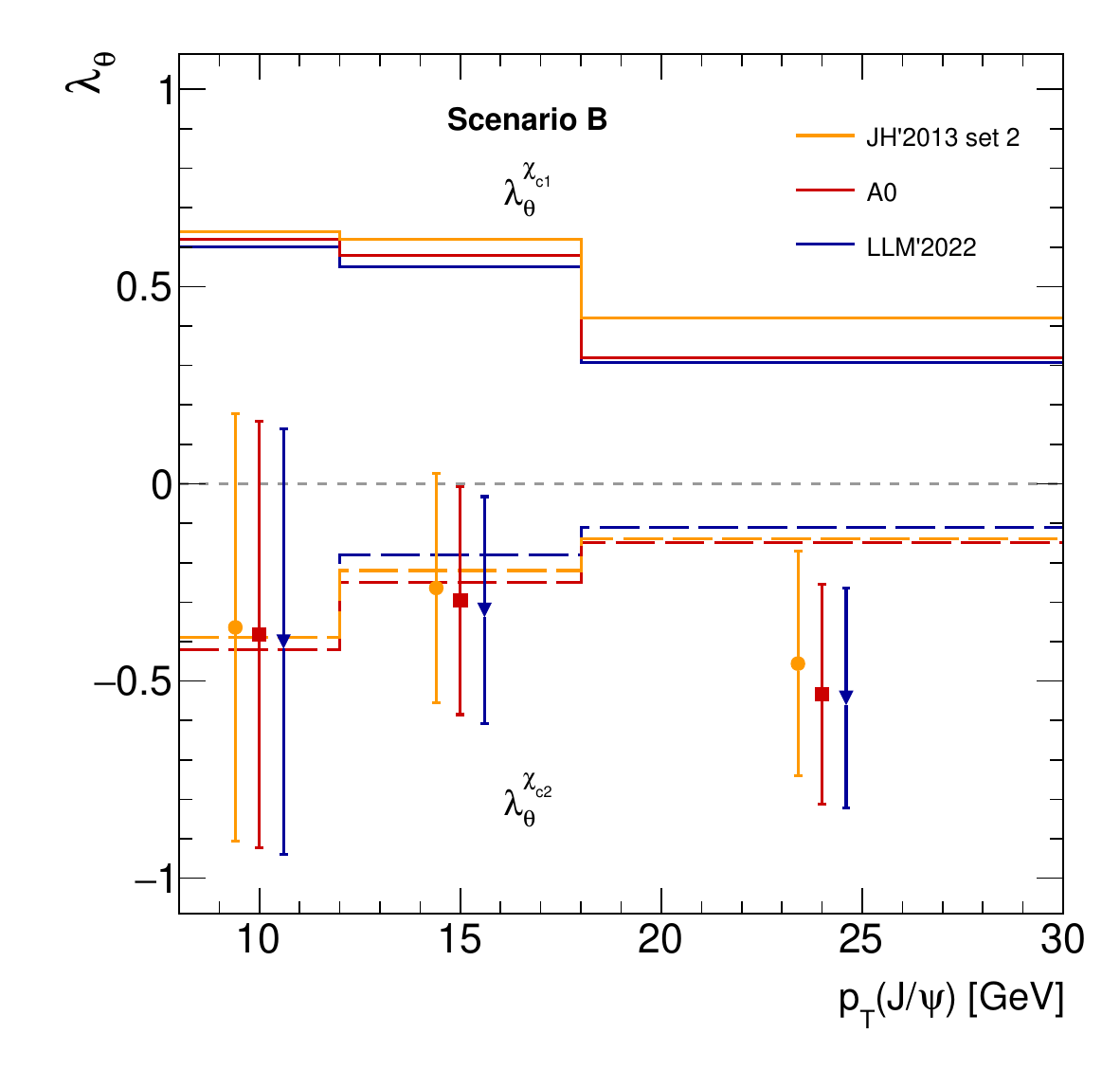}}\hfill
\caption{Polarization parameters $\lambda^{\chi_{c1}}_{\theta}$ and $\lambda^{\chi_{c2}}_{\theta}$ calculated at $\sqrt{s} = 8$ TeV 
in the rapidity region $|y^{J/\psi}| < 1.2$ for scenario A (left panel) and scenario B (right panel). 
``Experimental points'' represent the $\lambda^{\chi_{c2}}_{\theta}$ values obtained from the
experimentally established relations (\ref{eq:CMS_polar}).}
\label{fig:CMS_polar}
 \end{center}
\end{figure}

Next, the extracted LDMEs are employed to investigate the polarization 
of $\chi_{c1}$ and $\chi_{c2}$ mesons. We compare our predictions with the first results 
reported by the CMS Collaboration at $\sqrt s = 8$~TeV\cite{chic-CMS}, which have established 
certain correlations between the polarization parameters
$\lambda^{\chi_{c1}}_{\theta}$ and $\lambda^{\chi_{c2}}_{\theta}$.
The data were collected in the $J/\psi$ rapidity range $|y^{J/\psi}| < 1.2$ 
for three subdivisions of $p_T$, namely, 
$ 8 < p^{J/\psi}_{T} < 12$ GeV, $12 < p^{J/\psi}_{T} < 18$ GeV 
and $18 < p^{J/\psi}_{T} < 30$~GeV.
The muon angular distribution is conventionally parametrized (in the $J/\psi$ helicity frame) as
\begin{gather}
 {d\sigma(J/\psi \to \mu\mu)\over d\cos\theta^*\,d\phi^*}\sim{1\over 3+\lambda_\theta}
 \left(1+\lambda_\theta\cos^2\theta^*+\lambda_\phi\sin^2\theta^*\cos 2\phi^*
 +\lambda_{\theta\phi}\sin2\theta^*\cos\phi^* \right),
 \label{eq:fit}
\end{gather}
\noindent
where $\theta^*$ and $\phi^*$ are the positive muon polar and azimuthal angles,
so that the $\chi_{cJ}$ angular momentum is encoded 
in the polarization parameters $\lambda_\theta$, $\lambda_\phi$ and $\lambda_{\theta \phi}$.
A simple correlation between the 
$\lambda_{\theta}^{\chi_{c1}}$ and $\lambda_{\theta}^{\chi_{c2}}$
parameters was determined in the CMS analysis\cite{chic-CMS}:
 \begin{gather}
 \lambda^{\chi_{c2}}_{\theta} = (-0.94 + 0.90\lambda^{\chi_{c1}}_{\theta})\pm(0.51 + 0.05\lambda^{\chi_{c1}}_{\theta}),\; 8 < p^{J/\psi}_{T} < 12\; {\rm GeV}, \nonumber \\
 \lambda^{\chi_{c2}}_{\theta} = (-0.76 + 0.80\lambda^{\chi_{c1}}_{\theta})\pm(0.26 + 0.05\lambda^{\chi_{c1}}_{\theta}),\; 12 < p^{J/\psi}_{T} < 18\; {\rm GeV}, \nonumber \\
 \lambda^{\chi_{c2}}_{\theta} = (-0.78 + 0.77\lambda^{\chi_{c1}}_{\theta})\pm(0.26+0.06\lambda^{\chi_{c1}}_{\theta}),\; 18 < p^{J/\psi}_{T} < 30\; {\rm GeV}.
 \label{eq:CMS_polar}
 \end{gather}
\noindent 
To evaluate these parameters, we collect the events
simulated in the kinematic region defined by the CMS measurement\cite{chic-CMS}
and generate the decay muon angular distributions according to the production and
decay matrix elements. Then we can easily determine the polarization 
parameters $\lambda_{\theta}^{\chi_{c1}}$ and $\lambda_{\theta}^{\chi_{c2}}$
by applying a three-parametric fit based on~(\ref{eq:fit}).
The results of our calculations are presented in Fig.~\ref{fig:CMS_polar} 
for both merging scenarios A and B. It can be noted that 
there is fairly good agreement with the expected values of $\lambda^{\chi_{c2}}_{\theta}$ 
obtained at the fixed $\lambda^{\chi_{c1}}_{\theta}$ from (\ref{eq:CMS_polar}).
However, there is some discrepancy between the results obtained in scenarios A and B.
One can see that a better agreement with the CMS data is achieved within the scenario A.
The latter can be addressed to a different role of the LO and NLO$^*$ terms in these two schemes.
The $2 \to 2$ contribution provide lower polarization of the $^3P^{[1]}_{J}$ mesons
as compared to the $2 \to 1$ contribution.
 
In the merging scenario B, the NLO$^*$ contributions are more important (see Fig.~\ref{fig:NLO_comparison}), 
thus leading to a decrease in the overall $\chi_c$ polarization.
This effect is clearly seen in the behaviour of $\lambda^{\chi_{c1}}_{\theta}$.
The LO and NLO$^*$ contributions to the $^3S^{[8]}_1$ channel
additionally suppress the $\chi_c$ polarization with increasing transverse momentum.
Our predictions for the polarization parameters $\lambda^{\chi_{c1}}_{\theta}$ and 
$\lambda^{\chi_{c2}}_{\theta}$
are almost insensitive to the choice of TMD gluon densities.

Finally, we would like to reiterate that inclusion of NLO$^*$ 
terms in the $k_T$-factorization approach enables us to strictly adhere to the HQSS 
rules for both color singlet and color octet channels and describe simultaneously the 
available Tevatron and LHC data. 
Both the considered merging schemes provide a decent description of the data, 
althouh the present limitations in measuring the mesons transverse momenta do not
allow us to make a choice in favor of one of the scenarios.
 
\section{Conclusion} \indent

In the present paper we have considered inclusive $P$-wave charmonia production in proton-proton 
and proton-antiproton collisions at high energies
in the $k_T$-factorization QCD approach beyond the standard leading-order approximation.
For the first time we have included tree-level next-to-leading contributions to corresponding 
production cross sections and
proposed two scenarios which consistently merge the $2 \to 1$ and $2 \to 2$
off-shell production amplitudes. 
We have introduced and discussed a special conditions which necessary to avoid the 
well-known double counting problem when calculating the 
higher-order corrections in the $k_T$-factorization approach.

Using several CCFM-evolved gluon densities in a proton,
we have extracted long-distance matrix elements for $\chi_c$ mesons 
from a combined fit to available Tevatron and LHC data.
In contrast to previous leading order $k_T$-factorization calculations, our fits 
do not conflict with equalizing the color-singlet wave functions for $\chi_{c1}$
and $\chi_{c2}$ states.
The previously observed violation of the HQSS relations for $\chi_c$ mesons
can be explained by the absence of higher-order corrections in the corresponding 
${\cal O}(\alpha_s^2)$ off-shell production amplitudes.
Taking into account the NLO$^*$ contributions provides a way to restore the HQSS relations and to improve an overall description of the data, especially the data 
on the relative production rate $\sigma(\chi_{c2})/\sigma(\chi_{c1})$.
Moreover, this observable is found to be sensitive to the TMD gluon density in 
a proton, and the best description is achieved with JH'2013 set 2 gluon.
Finally, our predictions are in a good agreement with the first measurements of
the $\chi_c$ polarization at the LHC reported recently by the CMS Collaboration.

\section*{Acknowledgements} \indent

We thank G.I.~Lykasov, M.A.~Malyshev and H.~Jung for their interest, useful discussions
and important remarks. 
Our study was supported by the 
Russian Science Foundation under grant~22-22-00119.

\bibliography{chic22}

\end{document}